\documentclass[aps,pra,10pt,twocolumn]{revtex4-1}

\usepackage{xcolor}

\usepackage{dsfont}
\usepackage{amsfonts}
\usepackage{amsmath}
\usepackage{bm}
\usepackage{amssymb}
\usepackage{graphicx}
\usepackage{bm}%
\usepackage{braket}

\DeclareGraphicsExtensions{.eps,.ps,.pdf}
\newtheorem{theorem}{Theorem}
\newtheorem{lemma}{Lemma}
\newcommand{\diff}{\mathop{}\!\mathrm{d}}
\newcommand{\al}{\alpha}

\newcommand{\de}{\delta}
\newcommand{\la}{\lambda}

\newcommand{\si}{\sigma}

\newcommand{\pd}{\partial}

\newcommand{\BM}{\begin{displaymath}}

\newcommand{\EM}{\end{displaymath}}

\newcommand{\ie}{\hbox{\em i.e.{}}}

\def\nn{\nonumber}

\def \be  {\begin{equation}}
\def \eeq {\end{equation}}
\def \baa {\begin{eqnarray*}}
\def \eaa {\end{eqnarray*}}
\def \bb  {\begin{thebibliography}}
\def \eb  {\end{thebibliography}}

\def \lab #1 {\label{#1}}

\def \Tr {\text{Tr}}

\def \CP {{\cal P}}

\newcommand{\rhs}{\hbox{r.h.s.{}}}
\newcommand{\Rr}{\mathsf{R}}
\newcommand{\Hs}{\mathcal{H}}
\newcommand{\Bs}{\mathcal{B}}

\newcommand{\Id}{\mathds{1}}
\newcommand{\BHs}{\mathcal{B}}

\newcommand{\co}{\mathcal{C}}
\newcommand{\sco}{\mathbf{c}}

\newcommand{\da}{\dagger}

\newcommand{\Op}{\mathcal{O}}

\makeatletter
\g@addto@macro\bfseries{\boldmath}
\makeatother

\begin{document}
\widetext

\title{Majorana representation for mixed states}

\author{E.{} Serrano-Ens\'astiga}
\email{eduardo.serrano-ensastiga@uni-tuebingen.de}
\affiliation{Institut f\"ur Theoretische Physik \\
	Universit\"at T\"ubingen\\
	72076 T\"ubingen, Germany}

\author{D.{} Braun}
\email{daniel.braun@uni-tuebingen.de}	
\affiliation{Institut f\"ur Theoretische Physik \\
	Universit\"at T\"ubingen\\
	72076 T\"ubingen, Germany}
	
\begin{abstract}
	\noindent 
     We generalize the Majorana stellar representation of
  spin-$s$ pure states to mixed states, and in general
  to any hermitian operator, defining a bijective correspondence
  between three spaces: the spin density-matrices, a projective space
  of homogeneous polynomials of four variables, and a set of equivalence classes
  of points (constellations) on spheres of different radii. The
  representation behaves well under rotations by construction, and
  also under partial traces where the reduced density matrices inherit
  their constellation classes from the original state $\rho$.
    We express several concepts and operations related to density matrices in terms of the corresponding polynomials, 
such as the anticoherence criterion and the tensor representation of
  spin-$s$ states described in \cite{Gir.Bra.Bag.Bas.Mar:15}. 
\end{abstract}
\maketitle
\section{Introduction}
The Majorana stellar representation \cite{Majorana1932}  
enlightens, among other properties, an image of any spin-$s$ state,
and in consequence provides a glance of the (projective) 
Hilbert space $\Hs_s$ structure of the pure states. 
The representation defines a bijection between states 
$\ket{\psi} \in \Hs_s$ and $2s$ points (\emph{stars}) on the sphere
$S^2$, called the \emph{constellation} of $\ket{\psi}$,
$\co_{\psi}$. The spin coherent (SC) states \cite{Perelmov86,Rad:71},
which are the most  `classical' quantum states, have the simplest
constellations: all the stars point in the same
direction. In the opposite extreme, the most `quantum' states are
related to constellations spreading their stars over the unit sphere $S^2$, where the
`quantum' property can be measured in several ways, e.g., the quantumness
\cite{Giraud10,Boh.Bra.Gir:16}, anticoherence and {higher-order
  multipolar fluctuations
  \cite{Zimba06,Bag.Mar:17,Bag.Dam.Gir.Mar:17,Hoz.Kli.Bjo:13}}, and
states with maximal Wehrl-Lieb entropy \cite{Bae.Ing:14}. They have
important applications 
in quantum metrology, as they contain the most sensitive states
under small rotations for a known or unknown rotation axis
\cite{Kolenderski08,Bou.etal:17,Chr.Her:17,Gol.Jam:18}. The Hilbert space $\Hs_s$ as a whole can be seen as a stratified manifold foliated by the $SU(2)$ orbits of all the possible configurations of constellations \cite{Bengtsson17}. In addition, the Majorana constellation has been
useful in other applications, such as the classification of spinor 
Bose-Einstein-condensate phases
\cite{Bar.Tur.Dem:06,Bar.Tur.Dem:07,Mak.Suo:07} or the investigation
of the thermodynamical
limit in the Lipkin-Meshkov-Glick model
\cite{Rib.Vid.Mos:07,Rib.Vid.Mos:08}. This representation also plays a role in the
Atiyah mapping related to his conjecture on `configurations of points'
\cite{Ati:01}. 
Other  characterizations of quantum systems via  points on
a manifold are also commonly used. Examples include the use of zeroes of the Husimi
function, or zeroes of Haldane's trial wave function for the
fractional quantum Hall effect
\cite{Leb.Vor:90,Hal.Rey:85,Aro.Bha.Hal.Lit.Ram:88}.  

 Many representations and parametrizations have been found for pure
 and/or mixed spin states which also behave well under rotations
 \cite{Byr.Kha:03,Mak.Mes:10,Mak.Mes:10b,Aer.Bia:16,Bru.Mak.Mes.Pet:12,Ash.Sir:13}
 or Lorentz transformations  \cite{Gir.Bra.Bag.Bas.Mar:15}. Moreover,
 there are complete parametrizations of quantum states for small values of spin-$s$ \cite{giraud_parametrization_2012,Cab.Rem.Smo.Wal:15}.
 However, for the case of mixed states, none of them share all the
 properties of the standard Majorana representation for pure states
 as: bijection with a projective space of polynomials \footnote{We say the projective space of polynomials instead of the projectivization of a polynomial ring because we do not consider the multiplication operation of polynomials.}, bijection with a set of points
 in the physical space, and well--behavior under rotations. The
 Majorana representation for mixed states that we introduce in this paper
 has all these properties, with additional properties associated with
 the partial trace and the tensor product. While the bijection of mixed states with polynomials is new and studied here, the bijection with a set of points (constellations) in the physical space is described in a little known paper \cite{Ram.Rav:86}, and it uses the decomposition of the density matrix in irreducible representations of the $SU(2)$ group
 \footnote{This representation has been called the Multiaxial
   representation (MAR) in other works \cite{Ash.Sir:13,
     Sum.Sir.Hed.Bha:17}. However, we think that it is a confusing
   name because usually axes are unoriented objects while, as it is explained here and in the original work \cite{Ram.Rav:86}, the orientation of the axes provides relevant information to uniquely specify the state. If one does not consider the axes orientations (or subconstellations) as was done in \cite{Ash.Sir:13, Sum.Sir.Hed.Bha:17}, the mapping from the states to the MAR representation is not injective.}. We call it accordingly  the
 Ramachandran-Ravishankar representation, or $T$-rep for short. The $T$-rep associates to any density matrix a set of equivalence classes of constellations on spheres of different radii
 \cite{Ram.Rav:86}. The bijective correspondence between matrices and
   polynomials implies that the irreducible representations in both
   spaces are equal, and hence both of them end up with the same 
 stellar representation as the $T$-rep. There is another representation of mixed states close to the Majorana representation described in this paper, given by the tensor product of Pauli
 matrices projected in the fully symmetric sector. This tensor
 representation is described in \cite{Gir.Bra.Bag.Bas.Mar:15} and it
 has been helpful to study the problem of classicality of spin states
 \cite{boh.bra.gir:16.2}, to establish a relation between entanglement
 and the truncated moment problem \cite{Boh.Gir.Bra:17},  and to study
 genuinely entangled symmetric states \cite{Des.Gir.Mar:17}, among
 others works \cite{Boh.Bra.Gir:16,Mil.Bra.Gir:19}. The latter
 representation will be denoted as the $S$-rep. The connection between the $T$- and $S$-representations, not known until now,  is presented here.  

 The paper is organized as follows: In Sec. \ref{MajPol.rho} we
 present the Majorana polynomial for density matrices, the necessary
 elements to build it and the translation of the physical operations of interest to
this representation. In Section \ref{Sec.MajConst} we explain the Ramachandran-Ravishankar $T$-rep, \ie, the bijection between the mixed states and a set of equivalence classes of points on the physical space. The way we introduce this bijection is different from the pioneering paper \cite{Ram.Rav:86} but closer to our notation and definitions. In addition, we deduce the properties of the Majorana representation of mixed states with respect to partial traces, the polynomial expression of the anticoherence criterion, and the connection of the $S$- and $T$-reps. The relation between the Majorana polynomial for mixed states and the Husimi and P- quasiprobability distributions is explained in Sec. \ref{Sec.IV}.  We end the paper with some final comments in Sec.~\ref{Sec.Con}.
\section{Majorana polynomial for mixed states}
\label{MajPol.rho}
\subsection{The standard Majorana representation}
The Majorana stellar representation for pure spin-$s$ states \cite{Majorana1932,Chr.Guz.Ser:18}
associates one-by-one each point of the Hilbert space $\ket{\psi} \in \Hs_s$ with $N=2s$ points on the sphere $S^2$ that contains the full information of the state since
the real dimension of the projective Hilbert space, after taking out
the normalization and global phase factor of the state, is $
\dim(\Hs_s)=2N$. E. Majorana \cite{Majorana1932} defined this representation via a polynomial constructed with the expansion of the state $\ket{\psi}$ in the $S_z$-eigenbasis, $\ket{\psi} = \sum_{m=-s}^s \la_m \ket{s,m}$
\begin{equation}
p_{\psi}(Z) = \sum_{m=-s}^s (-1)^{s-m}   \sqrt{\binom{2s}{s-m}} \la_m Z^{s+m}  \, .
\label{first.pol}
\end{equation}
The complex roots of $p_{\psi}(Z)$ specify uniquely the polynomial and hence the state  $\ket{\psi}$ up to an irrelevant global complex factor. The polynomial $p_{\psi}(Z)$ has degree at most $N=2s$, and by a rule which be clarified later, its set of roots $\{ \zeta_k \}_{k}$ is always increased to $2s$ by adding the sufficient number of roots at the infinity.  The \emph{constellation} $\co_{\psi}$ of $\ket{\psi}$ is the set of points on $S^2$ called \emph{stars} obtained with the stereographic projection from the South Pole of the roots $\{ \zeta_k \}_{k=1}^{N}$, where the complex plane is situated in the $xy$-plane and the $\bm{x}$- and $\bm{y}$- axes are the real and imaginary axes, respectively. The stereographic projection maps the complex number $\zeta = \tan(\theta/2) e^{i \phi}$ to the point $\bm{n}$ on the sphere $S^2$ with polar and azimuthal angles $(\theta,  \, \phi)$. 

In order to generalize this polynomial to density matrices, we work with a similar representation defined by H. Bacry \cite{Bac:74} that associates to each state $\ket{\psi}$ a homogeneous polynomial of two variables, that it can be written as $p_{\psi}(z_1 ,z_2) \equiv \braket{-\bm{n}_B |\psi}$, where    
\begin{equation}
\bra{-\bm{n}_{B}} \equiv \sum_{m=-s}^s (-1)^{s-m}   \sqrt{\binom{2s}{s-m}} z_1^{s+m} z_2^{s-m} \bra{s, \, m} \, .
\label{BSC.state}
\end{equation}
Following the habit in quantum optics, we call $\ket{\bm{n}_B}$ the
\emph{Bargmann Spin Coherent} (BSC) state, which is proportional to
the Spin Coherent (SC) state pointing in the direction $\bm{n}$
associated with the complex number $z_1 / z_2$ via the stereographic
projection.  The latter polynomial, which we call also the Majorana
polynomial of $\ket{\psi}$ for simplicity, has the expression given by  
\begin{equation}
p_{\psi}(z_1 , \, z_2) = \sum_{m=-s}^s (-1)^{s-m} \sqrt{\binom{2s}{s-m}} \la_m z_1^{s+m} z_2^{s-m} 
\label{Maj.for} \, .
\end{equation}
In principle, one could work with the zeroes of the new polynomial
\eqref{Maj.for} and then one would associate to any state $\ket{\psi}$
an algebraic variety in $\mathds{C}^2$.  But this is more information than we need to specify a state and it is not easy to visualize. To avoid these complications, we use the fact that the polynomial \eqref{Maj.for} is homogeneous and hence the polynomial is fully factorizable
\begin{equation}
p_{\psi}(z_1 , \, z_2)  = \prod_{k=1}^{N} (z_1 \al_k - z_2 \beta_k) \, ,
\label{roots.pol}
\end{equation}
which implies that the polynomial is characterized by $N$ rays on $\mathds{C}^2$ $\{ (z_1, \, z_2) \in \mathds{C}^2 | z_1 \al_k - z_2 \beta_k =0 \}_{k=1}^N$, or equivalently, {by $N$ points $\{ \zeta_k= \beta_k / \al_k \}_{k=1}^{N}$ in the projective complex space $\mathds{P}(\mathds{C}^2) = \mathds{C} P^1$, defined by the set of (complex) rays in $\mathds{C}^2$ and isomorphic to the extended complex plane $ \in \mathds{C} \cup \{ \infty\}$}. The set $\{\zeta_k \}_{k=1}^N$ obtained here is equal to the set of roots defined by \eqref{first.pol} and hence the same constellation is obtained using the stereographic projection explained above. On the other hand, any spin-$s$ state is a fully symmetric state of $N$ constituent spin-$1/2$ states $\ket{\bm{n}_k}$
\begin{equation}
\ket{\psi} \propto \sum_{\pi \in S_N} \pi \left( \ket{\bm{n}_{1}}\otimes \dots \otimes \ket{\bm{n}_{N}} \right) \, ,
\label{sta.constituents}
\end{equation}
where the summation is over all the elements of the permutation group of $N$ elements $S_N$ and the spin-1/2 states are labeled by its respective Bloch vector $\bm{n}_k$. The definition of the Majorana polynomial implies that 
the stars of $\co_{\psi}$ are equal to the directions of the
constituents spin-$1/2$ of $\ket{\psi}$. In particular, the complex
number $\zeta_k =  \tan(\theta/2) e^{i \phi}$ with stereographic
projection $\bm{n}_k$ of angles $(\theta_k, \, \phi_k)$ is associated
to the constituent of $\ket{\psi}$, $\ket{\bm{n}_k} = \al_k \ket{1/2,
  1/2} + \beta_k \ket{1/2,-1/2} $ with $\beta_k / \al_k  = \zeta_k$
and $|\al_k|^2 + |\beta_k|^2=1$. To summarize, the Majorana stellar
representation defines bijective mappings among the Hilbert space
$\Hs_s$, the projective space of homogeneous bivariate polynomials of degree $N$ and the set of constellations on $S^2$ with
$N$ stars.

A transformation $\ket{\psi'}=U(\Rr)\ket{\psi}$ in $\Hs_s$ where the unitary transformation $U(\Rr)\equiv \exp(-i \bm{ e} \cdot \bm S \eta/\hbar)$ represents a rotation $\Rr \in SO(3)$ with rotation angle $\eta$ about the $\bm{e}$-axis of unit norm and angles $(\Theta , \, \Phi)$, and $\bm{S}=(S_x,S_y,S_z)$ is the vector of angular momentum operators, rigidly rotates 
the corresponding constellation $\co_{\psi} \subset  S^2$ with the same rotation $\Rr$. The roots of the Majorana polynomial of $\ket{\psi'} \in \Hs_s$ are $ \zeta'_k  = M(\zeta_k)$ for $k=1, \,\dots , \, N$  \cite{Bengtsson17} and
\begin{equation}
M(\zeta) = \frac{a \zeta - b }{b^* \zeta + a^*} \, 
\label{Mob.trans}
\end{equation}
is the M\"obius transformation associated to the rotation $\Rr$ with $a=\cos (\eta/2)- i \sin (\eta/2) \cos \Theta$ and $b = -i \sin (\eta/2) \sin \Theta e^{i \Phi}$  (\cite{Var.Mos.Khe:88}, p.27). The complex numbers $(a,b)$ with $|a|^2+|b|^2=1$ are called the Cayley-Klein parameters of a rotation $\Rr$. In polynomials, $p_{\psi'}(z_1, z_2) = p_{\psi} (z'_1,z'_2)$ where the new variables are
\begin{equation}
\left (
\begin{array}{c}
z_1' \\
z_2'
\end{array}
\right) = 
\left( 
\begin{array}{c c}
a^* & b
\\
-b^* & a
\end{array}
\right)
\left(
\begin{array}{c}
z_1
\\
z_2
\end{array}
\right) \, ,
\end{equation}
and the matrix 
\begin{equation}
 \left( 
\begin{array}{c c}
a^* & b
\\
-b^* & a
\end{array}
\right) \in SU(2) 
\end{equation}
is the projective matrix representation of the rotation $\Rr^{-1}$ and hence the matrix associated to the inverse of the M\"obius transformation \eqref{Mob.trans}.
The covariant transformation of the constellations implies that the point-group symmetry of $\co_{\psi}$ is the point-group symmetry of $\ket{\psi}$ under the respective unitary transformations
representing the symmetry operations. A similar statement holds true for Lorentz
symmetries, where invariants other than the shape of the constellation
become relevant, see \cite{Bac:74,Pen.Rin:90,Bengtsson17}. In this case, a Lorentz transformation is associated with a generic M\"{o}bius transformation 
\begin{equation}
M(\zeta) = \frac{a \zeta +b}{c \zeta +d} \, , \quad \quad \text{with } \, ad-bc =1 \, .
\end{equation}
\subsubsection*{ The derivative of the Majorana polynomial }
A not well-known result about the Majorana polynomial is about the physical meaning of its derivative. Here we explain it briefly following Sec. 2.6 of \cite{Lan:12}. The state $\ket{\psi}$ defined in Eq. \eqref{sta.constituents} becomes, after the contraction of the first of its constituents with, let us say, the spin-1/2 state pointing in the $\bm{z}$ direction $\ket{\bm{z}} = \ket{1/2,1/2}$, a state $\ket{\psi_{\bm{z}}} $ of spin $s'$ with $s'=s-1/2$ and proportional to 
\begin{equation}
\ket{\psi_{\bm{z}}}  \propto 
\sum_{k=1}^N \al_{k} 
\sum_{\pi \in S_{N-1}} \pi\left( \ket{\bm{n}_{1}}\otimes \dots \otimes  \widehat{\ket{\bm{n}_{k}}} \otimes \dots \otimes  \ket{\bm{n}_{N}} \right) \, ,
\end{equation}
where the hat means exclusion in the
expression. 
On the other hand, the derivative with respect to $z_1$ of $p_{\psi}(z_1 , \, z_2)$ given by \eqref{roots.pol} is equal to
\begin{align}
 \partial_{z_1} p_{\psi}(z_1 , \, z_2) = \sum_{k=1}^N \al_k \prod_{j \neq k} (\al_j z_1 - \beta_j z_2)
  \propto  \, p_{\psi_{\bm{z}}} (z_1 , \, z_2)  \, ,
\end{align}
\ie, {the partial derivative  $\partial_{z_1}p_{\psi}(z_1, \, z_2)$ of the Majorana polynomial of $\ket{\psi}$ is, up to an irrelevant global factor, the polynomial $p_{\psi_{\bm{z}}}(z_1 , \,z_2)$ of the spin-$s'$ state $\ket{\psi_{\bm{z}}}$}. This result can be generalized in each direction (not only along $\bm{z}$): for a
direction $\bm{m}$ with angles $(\theta,\phi)$ and its respective
spin-1/2 state $\ket{\bm{m}} = \cos (\theta/2) \ket{1/2,1/2} + \sin (\theta/2) e^{i
  \phi} \ket{1/2,-1/2}$, 
\begin{equation}
p_{\psi_{\bm{m}}}(z_1 , \,z_2) = \left( \cos (\theta/2) \partial_{z_1} - \sin (\theta/2) e^{-i \phi} \partial_{z_2} \right) p_{\psi}(z_1 , \, z_2) \, .
\end{equation} 
In particular $p_{\psi_{-\bm{z}}}(z_1 , \, z_2) = \partial_{z_2} p_{\psi}(z_1 , \, z_2)$ where the global phase factor is not relevant for the roots of the resulting polynomial and hence for the respective final state.
\subsection{Majorana polynomial for a density matrix and its partial traces}
We reviewed how to associate a bivariate homogeneous polynomial of degree $N$ to a spin-$s$ pure state $\ket{\psi}$, and how the contraction of one of its constituent spin-$1/2$ is associated with the derivative of its Majorana polynomial. Now, we want to generalize this result to spin-$s$ operators in $\Bs ( \Hs_s )$, in particular to mixed states. To achieve this goal, we apply the BSC states \eqref{BSC.state}
to a general density matrix $\rho$ 
from the left and right to obtain
\begin{align}
 p_{\rho}(z_a,z^a) =& \bra{-\bm{n}_B} \rho \ket{-\bm{n}_B} \, ,
 \label{ms.pol}
\end{align}
with $z^a \equiv z_a^*$ the conjugated complex variables of $z_a$ for $a=1,2$.
While kets transform covariantly under rotations via their respective irreps $U(\Rr)$, 
bras transform contravariantly \cite{BrinkSatchler68}, as well as the BSC states
variables, 
\begin{align}
\left (
\begin{array}{c}
z_1 \\
z_2
\end{array}
\right) \rightarrow &
\left( 
\begin{array}{c c}
a^* & b
\\
-b^* & a
\end{array}
\right)
\left(
\begin{array}{c}
z_1
\\
z_2
\end{array}
\right) 
\, ,
\nn
\\[2pt]
\left ( z^1 \, \, z^2 \right)
 \rightarrow &
\left( z^1 \, \, z^2 \right) 
\left( 
\begin{array}{c c}
a & -b
\\
b^* & a^*
\end{array}
\right)
\, .
\end{align} 
We consider the set of variables $(z) \equiv (z_a,z^a)$ for $a=1,2$ independent, \ie, $\partial_{a} z^b = \partial^{a} z_b =0$ and $\partial_{a} z_b = \partial^{a} z^b = \delta_{a b} $ where
$\pd_a = \partial_{z_a}$ and $\partial^a = \partial_{z^a}$, and partial
derivatives transform as the inverse of their variables. Let us mention that the Majorana polynomial \eqref{ms.pol} can be applied to a general operator $C$, and in this way we have defined a mapping between $\Bs (\Hs_s)$ 
and homogeneous polynomials $p_{C}(z) $ of degree $2N$  where each
monomial $z_1^{\al} z_2^{\beta} (z^1)^{\gamma} (z^2)^{\delta}$ of
$p_{C}(z)$ satisfies $\al+\beta=\gamma+\delta=N$.   The last property implies that 
\begin{equation}
z_a \partial_a p_{C}(z) = N p_{C}(z) \, , 
\quad 
z^a \partial^a p_{C}(z) = N p_{C}(z) \, ,
\end{equation}
for any operator $C$ and where from here and forth we use the Einstein sum convention for repeated indices. We denote the vector space of polynomials of four variables $(z_1,\, z_2 , \, z^1 , \, z^2)$ as $P^{(N,N)}(z)$ and $p_{C}(z)$ is called the \emph{Majorana polynomial of $C$}. The mapping between $\Bs (\Hs_s)$ and $P^{(N,N)}(z)$ is bijective. The space $P^{(N,N)}(z)$ has been used before in \cite{Koo:81} to calculate the Clebsch-Gordan coefficients in terms of the Hahn polynomials.

The Majorana polynomial for states $\rho$ presented here is related to the standard Majorana polynomial in the case of pure states. For instance, the polynomial of $\rho_{\psi} = \ket{\psi}\bra{\psi}$ is equal to
\begin{equation}
p_{\rho_{\psi}}(z) = p_{\psi}(z_a) \left(p_{\psi}(z_a)\right)^*
 \equiv p_{\psi}(z_a) \bar{p}_{\psi}(z^a) 
\, ,
\end{equation}
where {$\bar{p}_{\psi}(z^a)$} denotes that we only conjugate the coefficients of the polynomial. Let us give an example of the Majorana polynomial for spin-$1/2$. {The density matrix $\rho= \ket{\bm{n}} \bra{\bm{n}}$ has Majorana polynomial}
\begin{equation}
p_{\rho}(z) = (\al z_1 - \beta z_2) (\al^* z^1 -  \beta^* z^2) \, ,
\end{equation}
with $\al ={\cos (\theta/2)}$ and $\beta= \sin (\theta/2)e^{i
  \phi}$. {In particular, $p_{\rho}(z)= z^1 z_1$ and 
$p_{\rho}(z)= z^2 z_2$ for $\bm{n} = \pm \bm{z}$, respectively}. 
As we mentioned before, we can associate a polynomial to any operator. For instance, the Pauli matrices $\si_\mu$ for $\mu=x,y,z$ and the ladder operators $\si_{\pm} = \si_x \pm i \si_y$ have polynomials
\begin{align}
p_{0} (z)= &  z_a z^a \, , \quad & p_{x}(z) =& - z_1 z^2  -  z_2 z^1 \, ,
\nn
\\
p_{y} (z)= &  i \left( z_1 z^2  - z_2  z^1 \right)
\, , \quad & p_{z} (z) = & z_1 z^1 - z_2  z^2 \, ,
\nn
\\
p_{+} (z)= & -2 z_1 z^2 \, , \quad & p_{-}(z) = & -2z_2 z^1 \, ,
\label{Pau.pol}
\end{align}
with $p_{\mu}(z) \equiv p_{\si_{\mu}}(z) $ for $\mu=0, \, x, \, y, \, z , \, + , \, -$ and $\si_{0} = \Id_2$ is the $2\times 2$ identity matrix. The polynomial of the adjoint of an operator $A$,  $p_{A^{\dagger}}(z)$, is obtained interchanging $z_a \leftrightarrow z^a$ and conjugating the coefficients in $p_A(z)$. The polynomial of an Hermitian operator is invariant under this transformation. We can observe these properties in the {Pauli matrices and ladder operators}. 

According to the discussion of the previous subsection, the reduced density matrix $\rho_{s'}$ with $s'=s-1/2$ obtained by tracing the spin-$s$ state $\rho$ over a constituent spin-$1/2$, $\rho_{s'}= \Tr_1 (\rho)$, can be written as the application of the differential operator $\left( \partial_a \partial^a \right)$ to the Majorana polynomial of $\rho$. We define the \emph{partial trace} operator $L: P^{(N,N)}(z) \rightarrow P^{(N-1,N-1)} (z)$ as
\begin{equation}
L( p(z)) \equiv N^{-2} \partial_a \partial^a p(z)\, ,
\label{Norm.L}
\end{equation}
where, as we will see in Theorem \ref{Big.Theo}, the factor $N^{-2}$ guarantees that the trace of the operators is preserved. The operator $L$ is invariant under rotations due to the transformation laws of the partial derivatives. The application of the operator $L$ $2(s-k)$-times to $p_{\rho}(z)$ yields the associated polynomial of the spin-$k$ reduced density matrix $\rho_{k} = \Tr_{2(s-k)} (\rho)$,
\begin{equation}
p_{\rho_k}(z) = L^{2(s-k)}(p_{\rho}(z)) \, .
\end{equation} 
\subsection{Operations in $\Bs (\Hs_s) $ in terms of polynomials}
We are interested in making calculations in terms of polynomials, and here we briefly
deduce the most common operations in $\Bs (\Hs_s)$.  Let us start with
the trace of an operator $C$, given by the action of the partial trace
operator applied $N$ times 
\begin{equation}\label{traceop}
 p_{\Tr (C)}(z) = L^{N}(p_C(z)) =  (N!)^{-2} \left( \partial_a \partial^a \right)^N p_{C}(z) \, .
\end{equation}
{In particular, the identity matrix polynomial $p_{\Id}(z) =(z^a z_a)^{N}$ satisfies $p_{\Tr (\Id)}(z) = N+1$.}

Another basic operation in $\Bs (\Hs_s)$ is the calculation of an operator $C$ given by a product of operators
$C=DE$.  How to calculate it in terms of polynomials is the result of the following
\begin{lemma}
\label{Action.optoop}
 Let $C, \, D , \, E \in \Bs (\Hs_s ) $ such that $C=D E$ and with Majorana polynomials $p_{C}(z)$, $p_{D}(z)$, $p_E(z) \in  P^{(N,N)}(z)$, respectively. Then
\begin{align}
p_C(z)  &  
= (N!)^{-1} p_D(z_a , \partial_a) p_E(z_a , z^{a}) \, ,
\nn
\\
p_C(z)  & = (N!)^{-1} p_E( \partial^a , z^a) p_D(z_a , z^{a}) \, ,
\end{align}
where the order of the variables in each monomial of $p_D(z_a , \partial_a)$ and $
p_E( \partial^a , z^a)$ is such that the partial derivatives go to the
right of the monomial, to affect only the polynomial on the right.
\end{lemma}
 The result of lemma \ref{Action.optoop} can be applied iteratively for a product of several operators. In particular, an operator given by $C=DEF$ can be written in terms of differential operators acting on the polynomial $p_E(z_a , z^{a})$,
\begin{equation}
p_C (z) = (N!)^{-2}
 p_F( \partial^a , z^a) p_D( z_a , \partial_a) p_E(z_a , z^{a})
\, .
\end{equation}
For instance, the $X=\si_x$ channel $X \rho X$ has an output polynomial equal to
\begin{equation}
p_{X\rho X}(z) = \left( z^1 \pd^2 + z^2 \pd^1 \right) \left( z_1 \pd_2 + z_2 \pd_1  \right) p_{\rho}(z)  \, .
\end{equation}
 The combination of the trace and the product of operators has a simplified expression:
\begin{lemma}
\label{prod.op}
 Let $C, \, D \in \Bs (\Hs_s ) $ with Majorana polynomials $p_{C}(z)$, $p_{D}(z) \in  P^{(N,N)}(z)$. Then
\begin{equation}
\Tr \left( C D \right) = (N!)^{-2}  p_{C}(\pd^a , \pd_a) p_{D}(z_a , \,  z^a)  \, .
\end{equation}
In particular, if the operators are such that $\Tr (CD)=0$, hence $p_C(\pd^a,\pd_a) p_D(z)=0$.
\end{lemma}
The proofs of the previous lemmas can be found in Appendix \ref{proof.lemmas}. We end this section writing the expectation value of an operator $C$ in a pure state $\ket{\psi}$ with constellation $\co_{\psi}$. Using the polynomial of a pure state and Lemma \ref{prod.op}, we obtain that
\begin{align}
\braket{\psi | C| \psi} =& (N!)^{-2}  p_{\psi}(\partial^a) \bar{p}_{\psi} (\partial_a) p_{C}(z) 
\nn
\\
=& (N!)^{-2}
\pd^{\bm{n}_{N}} \dots \pd^{\bm{n}_{1} } \pd_{\bm{n}_{N}} \dots \pd_{\bm{n}_{1}}   p_{C}(z) \, ,
\label{inne.rho}
\end{align}
where $\{ \bm{n}_k \}_{k}$ is the set of stars of  $\co_{\psi}$ with angles $(\theta_k , \, \phi_k)$ and
\begin{align}
\pd^{\bm{n}_k} = & \cos (\theta_k/2) \partial_{z^1} - \sin (\theta_k/2) e^{i \phi_k} \partial_{z^2} \, ,
\nn
\\
\pd_{\bm{n}_k} = & \cos (\theta_k/2) \partial_{z_1} - \sin (\theta_k/2) e^{-i \phi_k} \partial_{z_2} \, .
\end{align}
 The positive semidefinite condition of a state $\rho$ can be written as the condition that \eqref{inne.rho} is non-negative for any $N$ points $\{ \bm{n}_k \}$. As an extra result, we obtain that the only polynomials $p(z) \in P^{(N,N)}(z)$ such that $p(z_a,\pd_a) p(z_a,z^a) =N! \, p(z_a, z^a)$ are the polynomials associated with a pure state $p(z)=p_{\psi}(z_a) \bar{p}_{\psi}(z^a)$, \ie, polynomials that are factorizable with respect to the variables $(z_a)$ and $(z^a)$.
\section{Constellations for mixed states}
\label{Sec.MajConst}
 The Majorana representation for pure states allows us to visualize
 any state $\ket{\psi}$ via the stereographic projection of the roots
 of the polynomial $p_{\psi}(z)$. For the case of mixed states $\rho$,
 the equation $p_{\rho}(z_a , z^a)=0$ defines an algebraic variety on
 $\mathds{C}^4$, or $\mathds{C}^2$ taking into account that $z^a =
 z_a^*$. {The algebraic variety is not, in general, the product of a set
 of rays, and hence its projection in the extended complex plane is
 not necessarily a set of finite points}. A extreme case is given by the maximally mixed
 state $\rho^* = (2s+1)^{-1} \Id $, with $p_{\rho^*}(z) = (2s+1)^{-1}
 (z^az_a)^{2s} $ and hence the equation to fulfill is $p_{\rho^*}
 (z)\propto \left( |\zeta |^2 +1 \right)^{2s} = 0$ with $\zeta= z_1 /
z_2$. Instead of working with the zeroes of the full Majorana
polynomial $p_{\rho}(z)$, and in order to represent a state with a
finite set of points, we work with the irreducible representations (irrep) of
$SU(2)$ in $P^{(N,N)}(z)$, or equivalently, in $\Bs (\Hs_s)$. The $SU(2)$-irreps of $\Bs (\Hs_s)$ are spanned by the well-known
(multipolar) tensor operators $\{T_{\si \mu}^{(s)} | 0
\leq \si \leq 2s, |\mu| \leq \si \}$, and their use to associate to any mixed state a set of points in the physical space was discovered by Ramachandran and Ravishankar \cite{Ram.Rav:86}, leading to what we called the Ramachandran-Ravishankar representation or
\emph{T-rep} for short. In order to make the paper self-contained and for a better understanding of the next sections, we explain the $T$-rep in terms of Majorana constellations. The $T$-rep has been used  recently in Quantum Information
\cite{Ash.Sir:13,Sum.Sir.Hed.Bha:17}. 
\subsection{$T$-representation}
A tensor  operator $T_{\si \mu}^{(s)} : \Hs_s \rightarrow \Hs_s$ \cite{Fan:53,BrinkSatchler68,Var.Mos.Khe:88}
of rank  $\sigma$ is an element of a
 set of linear operators $\{T^{(s)}_{\sigma \mu} \}_{\mu=-\sigma
 }^{\si} $ that transforms under 
 a unitary transformation $U(\Rr)$ representing a rotation $\Rr \in
 SO(3)$ according to an \emph{irrep} 
 $D^{(\sigma)}(\Rr)$  of $SO(3)$ (or equivalent, of $SU(2)$),  
\begin{equation}
\label{prop.sh}
U(\Rr) T^{(s)}_{\sigma \mu} U^{-1}(\Rr) = \sum\limits_{\mu'=-\sigma}^{\sigma} D_{ \mu' \mu}^{(\sigma)}(\Rr) T^{(s)}_{\sigma \mu'} \, ,
\end{equation}
where $D^{(\si)}_{ \mu' \mu} (\Rr) \equiv \braket{\si , \, \mu' |
    e^{-i \al S_z} e^{-i \beta S_y} e^{-i \gamma S_z} | \si , \, \mu}$
  is the Wigner D-matrix \cite{Var.Mos.Khe:88} of a rotation $\Rr$
  with Euler angles $(\al, \, \beta , \, \gamma)$, and
$\sigma=0,1,2,\ldots$ labels the irrep. The explicit expression of
$T_{\sigma \mu}^{(s)}$ can be given in terms of the Clebsch-Gordan
coefficients $C_{j_1 m_1 j_2 m_2}^{j m}$, 
\begin{equation}
\label{decomp.TensOp}
T_{\sigma \mu}^{(s)}
=
\sum_{m,m'=-s}^s (-1)^{s-m'} C_{sm,s-m'}^{\si \mu}\ket{s,m}\bra{s,m'} 
\, .
\end{equation}
From now on, we omit the super index $(s)$ when there is no possible confusion.
 It is easy to deduce that $0\leq \sigma \leq 2s$, $|\mu|\leq \si$ and the following
properties: 
\begin{equation}
\Tr ( T_{\sigma_1 \mu_1}^{\dagger} T_{\sigma_2 \mu_2} ) = \delta_{\sigma_1 \sigma_2} \delta_{\mu_1 \mu_2} \, , \quad T_{\sigma \mu}^{\dagger} = (-1)^{\mu} T_{\sigma-\mu} \, .
\label{hermi.tens.op}
\end{equation} 
The set $\{T_{\sigma \mu} : 0 \leq \sigma \leq 2s , \, -\sigma \leq
\mu \leq \sigma \}$ forms hence an orthonormal  basis over the complex numbers for the complex square matrices of order $N+1$ satisfying (\ref{prop.sh}). In other
words, the set of $T_{\sigma \mu}$ is the matrix analogue of the spherical
harmonic functions $Y_{lm} (\theta, \, \phi )$, which span the space
of real-valued functions on the sphere $f(\theta , \phi)$. A density
matrix $\rho \in  \BHs (\Hs_s)$ has then a block decomposition in the $T_{\sigma \mu} $ basis
\begin{equation}
\rho = \sum_{\si=0}^{2s} \bm{\rho}_{\si} \cdot \bm{T}_{\si} \, ,
\label{bloc.decomp}
\end{equation}
where $\bm{\rho}_{\si}= (\rho_{\si \si} , \dots ,\rho_{\si -\si}) \in \mathds{C}^{2\si+1}$ with
$\rho_{\si \mu} = \Tr( \rho \, T^{\da}_{\si \mu})$, $\bm{T}_{\si}=
(T_{\si \si} , \dots , T_{\si , -\si})$ is a vector of matrices, and
the dot product is short for $\sum_{\mu=-\sigma}^\sigma\rho_{\si \mu}T_{\si \mu} $.
Each vector
$\bm{\rho}_{\si}$ can be associated to a constellation \emph{\`a la
  Majorana} \eqref{Maj.for} consisting of $2\si$ points on $S^2$
obtained with the stereographic projection of the complex roots of the polynomial $p_{\rho}^{(\si)}(z_1=\zeta,z_2=1)$ defined as 
\begin{equation}
p^{(\si)}_{\rho} (\zeta) = \sum_{\mu =-\si}^{\si}
(-1)^{\si-\mu}\sqrt{\binom{2\si}{\si-\mu}}
\rho_{\si \mu} \,  \zeta^{\si+\mu}
\,.
\label{pjk.1} 
\end{equation}
The respective constellation is denoted as $\co_{\rho}^{(\si)}$ or
$\co^{(\si)}$ when there is no possible confusion. The vector
$\bm{\rho}_0=(\rho_{00})$ does not have an associated constellation
and its value is fixed to $\rho_{00} = (2s+1)^{-1/2}$ by $\Tr \rho=1$. On the other hand, the hermiticity condition implies that  
\begin{equation}
\rho_{\si \mu}= (-1)^{\mu} \rho_{\sigma -\mu}^* \, , \quad \text{for all } \, |\mu | \leq \si \, ,
\label{her.ms}
\end{equation}
 which in turn implies that every constellation $\co^{(\si)}$ has antipodal
 symmetry. For a proof it is enough to show that if $\zeta=\xi$ is a root of
 $p^{(\si)}_{\rho}(\zeta)$, the corresponding antipodal complex number
 $\xi^A \equiv -1/\xi^{*}$ is also a root: 
\begin{align}
p_{\rho}^{(\si)}(\xi) =& \sum_{\mu}  (-1)^{\si-\mu}\sqrt{\binom{2\si}{\si-\mu}}
\rho_{\si \mu} \,  \xi^{\si+\mu}
\nonumber
\\
 =& \xi^{2\si} \left( \sum_{\mu}  (-1)^{\si-2\mu}\sqrt{\binom{2\si}{\si+\mu}}
\rho_{\si \, \mu} \,  \xi^{* -\si-\mu} \right)^*
\nonumber
\\
=& (-1)^{\si} \xi^{2 \sigma}  \left( p_{\rho}^{(\si)}(\xi^A)  \right)^* \, ,
\end{align}
where in the second equality we use \eqref{her.ms}. Hence, the proof is done for any root $\xi \neq 0$ but the statement also holds in the case of $\xi=0$
and its corresponding antipodal point $\xi= \infty$: Let us suppose that
the constant term in $p^{(\sigma)}_{\rho}(\zeta)$ is zero, and hence
there is a root $\xi=0$. The hermiticity property \eqref{her.ms} implies
that the coefficient of the highest exponent $\zeta^{2\sigma} $ is also
zero, implying that $p^{(\sigma)}_{\rho}(\zeta)$ has an extra root at
infinity. Hence, the roots at zero and infinity come also in pairs.

{The standard Majorana representation associates to each pure spin-$s$ state a unique polynomial up to a global factor, which does not change its roots and is of
no concern as the state can always be assumed normalized and the global phase is
irrelevant.} But now a pre-factor of a polynomial
$p^{(\sigma)}_{\rho}(\zeta)$ is a {\em relative} factor that carries
important information about the relative weights and phases of
different irreps contained in the state. Hence the set of
constellations of a state is not sufficient yet to specify the state
uniquely: Two states $\rho$ and $\rho'$ with the same constellation
$\co_{\rho}^{(\si)}=\co_{\rho'}^{(\si)}$ have the same vector
$\bm{\rho}_{\si}$ only up to arbitrary complex weights $w_{\si} e^{i
  \phi_{\si}}$ (in polar coordinates) that need to be given in
addition to the constellations in order to fully specify the state.  
In order to do so, we specify for each 
constellation $\co^{(\si)}$ the absolute value of the
weight with respect to a vector
$\tilde{\bm{\rho}}_{\si}$ with unit norm, $\bm\rho_\sigma=w_\sigma\tilde{\bm\rho}_\sigma$. The state $\rho$ can then be written as 
\begin{equation}
\rho = \frac{\mathds{1}}{2s+1} + \sum_{\si=1}^{2s} w_{\si}  \tilde{\bm{\rho}}_{\si} \cdot \bm{T}_{\si} \,,
\label{Ste.mix}
\end{equation}
with
\begin{equation}
\label{rhoT}
 w_{\si} = \left( \sum_{\mu=-\si}^{\si} \rho_{\si \mu}
  \rho_{\si \mu}^* \right)^{1/2}   \, ,
\end{equation}
and in particular $w_0 = \rho_{00} = (2s+1)^{-1/2}$. For the phase factor $e^{i \phi_{\si}}$, one could define 
a ``gauge''  for each $\si$-block, \ie, for each 
constellation $\co^{(\si)}$ one could specify a particular normalized vector $\tilde{\bm{\rho}}_{\si}^g$ that works as a reference to the phase factor $\tilde{\bm\rho}_\sigma = e^{i \phi_{\si}} \tilde{\bm{\rho}}_{\si}^g $. 
A disadvantage of fixing the gauge in this way is that
under rotations, the phase factor can have non-trivial transformation
laws. In fact, we know that a generic spin state may pick up an extra global phase after it traces a closed trajectory in the quantum states space by a sequence of rotations, which is the so-called geometric phase \cite{Chr.Jam:04}. The best way to
handle the phase factor is the following: First, let us remark that two normalized $(2\si+1)-$vectors $\tilde{\bm{\rho}}_{\si}$ and $\tilde{\bm{\rho}}'_{\si}$ that represent the $\si$-block of a physical state with equal constellation $\co^{(\si)}$, can differ only by a phase factor $\tilde{\bm{\rho}}_{\si} = e^{i \phi} \tilde{\bm{\rho}}'_{\si}$ with $e^{i \phi}= \pm 1$, otherwise one of the vectors does not satisfy the hermiticity condition \eqref{her.ms}. On the other hand, again by the hermiticity condition, the constellation $\co^{(\si)}$ defined by the $(2\si+1)-$vector $\tilde{\bm{\rho}}_{\si}$ has antipodal symmetry. This implies that there exists $\si$ stars $\sco \equiv (\bm{n}_1 , \, \dots , \, \bm{n}_{\si})$ in $\co^{(\si)}$ such that 
\begin{equation}
\{ \sco \} \cup \{ -\sco \} = \co^{(\si)}
\, ,
\label{subcons.cond}
\end{equation} 
 with $-\sco \equiv (-\bm{n}_1 , \, \dots , \, -\bm{n}_{\si})$ and where $\sco$ is a tuple and $\{ \sco\}$ its respective unordered set, and the same for $-\sco$ and $\{-\sco \}$. In general, the tuple $\sco$ that satisfies \eqref{subcons.cond} is not unique. The other choices can be written with respect to $\sco$ inverting the direction of some of its stars $\bm{\gamma}  \sco \equiv ( \gamma_1 \bm{n}_1 , \, \dots , \, \gamma_{\si} \bm{n}_{\si})$ with $\gamma_k = 1$ or $-1$.
 For simplicity, we refer to the unordered set $\{ \bm{\gamma} \sco \}$ with the same symbol as the tuple $\bm{\gamma} \sco$, and we call it a \emph{subconstellation} of $\co^{(\si)}$. Now, we can define a spin-$\si$ state for each subconstellation $ \bm{\gamma} \sco$ given by the projected bipartite state $ \CP_{\si} \ket{\phi , \, \phi^A} \equiv \CP_{\si} ( \ket{\phi} \otimes \ket{\phi^A} )$, with $\CP_{\si}$ the projection operator in the fully symmetric subspace of spin-$\si$ states, $\ket{\phi}$ the spin-$\si/2$ state with Majorana constellation $ \bm{\gamma}\sco $, and $\ket{\phi^A} \equiv A\ket{\phi}$, where $A$ is the \emph{time-reversal operator} defined by 
\begin{equation}
A\ket{\phi} \equiv \sum_{m} (-1)^{s+m} \la^*_{-m}\ket{s, \, m}, \, \, \text{for} \, 
\ket{\phi} = \sum_{m} \la_{m}\ket{s, \, m}.
\label{antipodal.def}
\end{equation}
We also call $A$ the \emph{antipodal operator} because the constellation of $\ket{\phi^A}$ is $ -\bm{\gamma} \sco $. The projector operator $\CP_{\si}$ is a function with domain the states space of $2\si$ spins-$1/2$ $(\Hs_{1/2})^{2\si}$ and image the set spanned by the symmetric Dicke states $\ket{D_{2\si}^{(m)}}$
\begin{equation}
\label{Dicke.state}
\ket{D_{2\si}^{(m)}} = K \sum_{\pi } \pi ( \underbrace{\ket{\bm{z}}\otimes \dots \otimes \ket{\bm{z}}}_{2\si-m} \otimes \underbrace{\ket{-\bm{z}} \otimes \dots \otimes \ket{-\bm{z}}}_{m}  ) \, ,
\end{equation}
with $K=\binom{2\si}{\si-m}^{-1/2}$ and where the sum runs over the permutations of $2\si$ objects of two types, with $2\si-m$ of the first type and $m$ of the second one. The symmetric Dicke states coincide with the $S_z$-eigenbasis $\ket{\si , \mu}$. From now on, we consider the projector operators restricted to its image $\CP_{\si} : (\Hs_{1/2})^{2\si} \rightarrow \Hs_{\si}$, and the direction of its action, left or right, is implicitly given in the equation. The expansion of the state $\CP_{\si} \ket{\phi , \, \phi^A}$ in the $S_z$-eigenbasis $\ket{\si , \mu}$ constitutes a $(2\si+1)-$vector that satisfies the hermiticity condition \eqref{her.ms} and produces the same constellation of $\tilde{\bm{\rho}}_{\si}$, $\co^{(\si)}$, 
\begin{equation}
 \CP_{\si} \ket{\phi , \, \phi^A} \propto \ket{ \pm \bm{n}_1 , \, \dots , \, \pm \bm{n}_{\si}  } \, ,
\label{sta.class}
\end{equation}
with $\ket{ \pm \bm{n}_1 , \, \dots , \, \pm \bm{n}_{\si}}$ the spin-$\si$ state with constellation $\co^{(\si)}$. Moreover, if one changes $\ket{\phi}$ by a phase factor $e^{i \delta} \ket{\phi}$, the coefficients of \eqref{sta.class} are invariant. On the other hand, if one turns the direction $\gamma_k \rightarrow -\gamma_k$ of a star of $\bm{\gamma} \sco$, the state \eqref{sta.class} remains equal times a global factor $-1$ because $ \CP_1 \left( \ket{\bm{n} } \otimes \ket{\bm{n}^A} \right) = -
\CP_1 \left( \ket{\bm{n}^A } \otimes \ket{ (\bm{n}^A )^A} \right) $ and the states $\ket{\phi}$ and $\ket{\phi^A}$ are fully symmetric $\si/2$-states. The last result suggests us to split the subconstellations $\sco \subset \co_{\rho}^{(\si)}$ satisfying \eqref{subcons.cond} into two equivalence classes, where two subconstellations are equivalent if they differ by an even number of stars. Both equivalence classes can be defined with respect to a particular subconstellation $\sco = \{ \bm{n}_k \}_{k}$ 
\begin{align}
& \left\{  \bm{\gamma} \bm{c} \subset \co^{(\si)} \Big|  \gamma_k= 1 \text{ or } -1 \, \text{ and }   \prod_{k=1}^{\si} \gamma_k =+ 1  \right\} \, ,
\nn
\\
& \left\{ \bm{\gamma} \bm{c} \subset \co^{(\si)} \Big|  \gamma_k= 1 \text{ or } -1 \, \text{ and }   \prod_{k=1}^{\si} \gamma_k =- 1   \right\} \, .
\label{class.cons}
\end{align}
Any element of any class produces the same state \eqref{sta.class},
but only elements of the same class produce the same
$(2\si+1)$-vector, \ie, the same state and the same phase
factor of the $(2\si+1)$-vector. In particular, for a state $\rho$ and for each $\si=1, \dots , 2s$, the vector 
$\tilde{\bm{\rho}}_{\si}$ belongs to one of these classes, with the respective constellations of $\rho$.  We denote the belonging \emph{subconstellation class} of $\tilde{\bm{\rho}}_{\si}$ of the state $\rho$ by $[\sco_{\rho}^{(\si)}] $,  
  with $\sco$ a representative element of the class. The components of $\tilde{\bm{\rho}}_{\si}$ can be written as
\begin{equation}
\tilde{\rho}_{\si \mu} = N_{\phi} \bra{ \si , \mu } \CP_{\si} \ket{ \phi, \phi^A}  \, ,
\label{cons.sector}
\end{equation}
where $\ket{\phi} = \ket{\bm{n}_1 , \dots , \bm{n}_{\si}}$ is a state with constellation lying in the class $[\sco^{  (\si)}_{\rho} ]$,  and $N_{\phi}$ a positive factor that guarantees $\tilde{\bm{\rho}}_{\si }$ is a normalized vector, namely
\begin{equation}
N_{\phi} = \frac{\left| \braket{\pm \bm{n}_1 , \, \dots , \, \pm \bm{n}_{\si} | \bm{n}_1  \otimes -\bm{n}_1 \otimes \dots \otimes \bm{n}_{\si} \otimes  -\bm{n}_{\si}}\right|}{ \left| \braket{ \bm{n}_1 \, , \dots \, , \bm{n}_{\si} | \bm{n}_1 \otimes \dots \otimes \bm{n}_{\si}}\right|^2}  \, ,
\label{Norm.factor}
\end{equation}
with $\ket{\bm{n}_1 \otimes \dots \otimes \bm{n}_{\si}} =
\ket{\bm{n}_1} \otimes \dots \otimes \ket{\bm{n}_{\si}}$. The scalar product \eqref{cons.sector} is given by 
\begin{equation}
  \label{eq:42b}
\bra{ \si , \mu } \CP_{\si} \ket{ \phi, \phi^A} =\braket{\phi | T_{\si \mu}^{ \da (\si/2)}  | \phi } \, .
\end{equation}
We conclude that any density matrix $\rho$ is uniquely specified
through $2s$ subconstellation classes $[\sco^{(\si)}]$ and $2s$
non-negative real numbers $w_{\si}$ considered as the radii of the
spheres where each subconstellation class lies, respectively. The
(continuous) degrees of freedom that parametrize the subconstellation
classes  $[\sco^{ (\si)}]$ are  $2\si$, and therefore the number of
free continuous parameters in $\{ w_{\si}, [\sco^{ (\si)}]
\}_{\si=1}^{2s}$ is $4s^2+4s$, the same as the number of real degrees
of freedom of the mixed states $\rho\in\mathcal B(\Hs_s)$.\\

The correspondence also works for any Hermitian operator $H$, where in this case the component $\Tr (H T_{00} )$ is not fixed. In addition, the correspondence between physical states $\rho$ and the set $\{ w_{\si} , [\sco^{(\si)}] \}$ is bijective. The parameters domain is restricted by the positive semidefinite condition $\braket{\psi|\rho|\psi}\geq 0$ for all $\ket{\psi}\in \Hs_s$, which with the unit trace condition $\Tr \rho=1$ implies that all eigenvalues of $\rho$ are in [0,1]. This condition is considerably more complicated to impose compared to hermiticity and unit trace. One necessary condition for positivity is that $\Tr \rho^2 \leq 1$, which gives an inequality independent of the subconstellation classes, 
\begin{equation}
\sum_{\si=1}^{2s} w_{\si}^2 \leq \frac{2s}{2s+1} \, .
\label{rad.wei}
\end{equation}
However, the positivity condition leads in general to a dependence of
the allowed range of the radii $w_{\si}$ on the classes $[ \sco^{(\si)} ]$.  

As an example, let us consider the $s=1/2$ case. 
Any vector $\bm{r}=(r_x, \, r_y ,\, r_z)$ with norm $r \leq 1$ is associated with a density matrix $\rho$ 
\begin{equation}
\rho = \frac{1}{2}( \mathds{1} + \bm{r}\cdot \bm{\si}) \, ,
\end{equation}
where $\bm \si = (\si_x \, ,\si_y \, ,\si_z)$ are the Pauli
matrices and $\bm{r}$ is called the Bloch vector of $\rho$. For a general spin value $s>0$, the necessary tensor operators with $\si=1$ are \cite{Aga:13}
\begin{align}
\label{TO.1}
T^{(s)}_{10}= & \left(\frac{3}{s(s+1)(2s+1)} \right)^{1/2} S_z \, ,
\\
T^{(s)}_{1,\pm1}=& \mp \left(\frac{3}{2s(s+1)(2s+1)} \right)^{1/2} S_{\pm}  \, .
\end{align}
The state $\rho$ written in the $T$-rep has a unique vector $\bm{\rho}^{(1)}$ equal to
\begin{equation}
\bm{\rho}^{(1)} = \frac{1}{2} \left(-r_x + i r_y , \sqrt{2} r_z , r_x + i r_y \right) \, .
\label{rho.1}
\end{equation}
The radius $w_1$ is obtained after normalization of the vector \eqref{rho.1}, yielding that
  $w_1=r/\sqrt{2}$. The condition \eqref{rad.wei} imposes that $
  \sqrt{2} w_1 = r \leq 1$, while the constellation $\co^{(1)}$ is
specified with the roots $\{\zeta\}$ of the  Majorana polynomial associated to the vector
$\bm{\rho}^{(1)}$. It is obtained that $\zeta_1= \tan (\theta /2 )e^{i
  \phi}$ and $\zeta_2=\zeta_1^A$ the antipodal complex number of $\zeta_1$,
with $(\theta, \phi)$ the spherical angles of the vector $\bm
r$. Therefore, the stars of $\co^{(1)}$ point in the parallel and
anti-parallel directions of the Bloch vector $\pm \bm{r} $. Lastly,
the subconstellation classes $[\sco^{ (1)}]$ are $[\bm{r}]$ and $[-\bm{r}]$, where each class has a unique element. We deduce the class 
to which the state \eqref{rho.1} belongs by comparing with the coefficients of the state $\CP_{1} \left(
  \ket{\bm{r}} \otimes \ket{\bm{r}^A} \right)$. For instance, with the parametrization $\ket{\bm{r}} = \cos (\theta/2)\ket{1/2, \,1/2} +  
\sin(\theta/2) e^{i \phi} \ket{1/2, \,-1/2}$, its coefficients in the $\ket{s=1,m}$ basis are
\begin{equation}
\frac{1}{2}  \left( -\sin(\theta) e^{-i \phi} \, , \sqrt{2} \cos \theta \, , \sin(\theta) e^{i \phi} \right) \, ,
\end{equation}
which are the same coefficients as in \eqref{rho.1} for $\bm{\rho}_1$  in spherical coordinates and hence its class is $[\bm{r}]$. 
Conversely, given a particular set $\{ w_1 , [\sco^{ (1)}]\}$, we can obtain the respective density matrix, \ie, the Bloch vector $\bm{r}$. We remark that states $\rho$ may differ only by some subconstellation classes $[\sco_{\rho}^{(\si)}]$, as in our $s=1/2$ example where the states with Bloch vector $\pm\bm{r}$ have the same constellation $\co^{(1)}$ but different class $[\pm \bm{r}]$. We can generalize the relation between antipodal states $\rho$ and $\rho^A = A\rho A^{\dagger}$ using the fact that $A$ is anti-unitary and $A^2 \ket{\psi} = (-1)^{2s} \, \ket{\psi}$,
\begin{equation}
\rho^{A} = A\rho A^{\dagger}= \sum_{\si=0}^{2s}  (-1)^{\si} \bm \rho_{\si} \cdot \bm{T}_{\si} \, .
\label{rel.anti.states}
\end{equation}
Therefore, the states $\rho$ and $\rho^A$ differ only by the subconstellation classes $[\sco^{(\si)}]$ of $\si$ odd.

\begin{figure}
 \scalebox{0.5}{\includegraphics{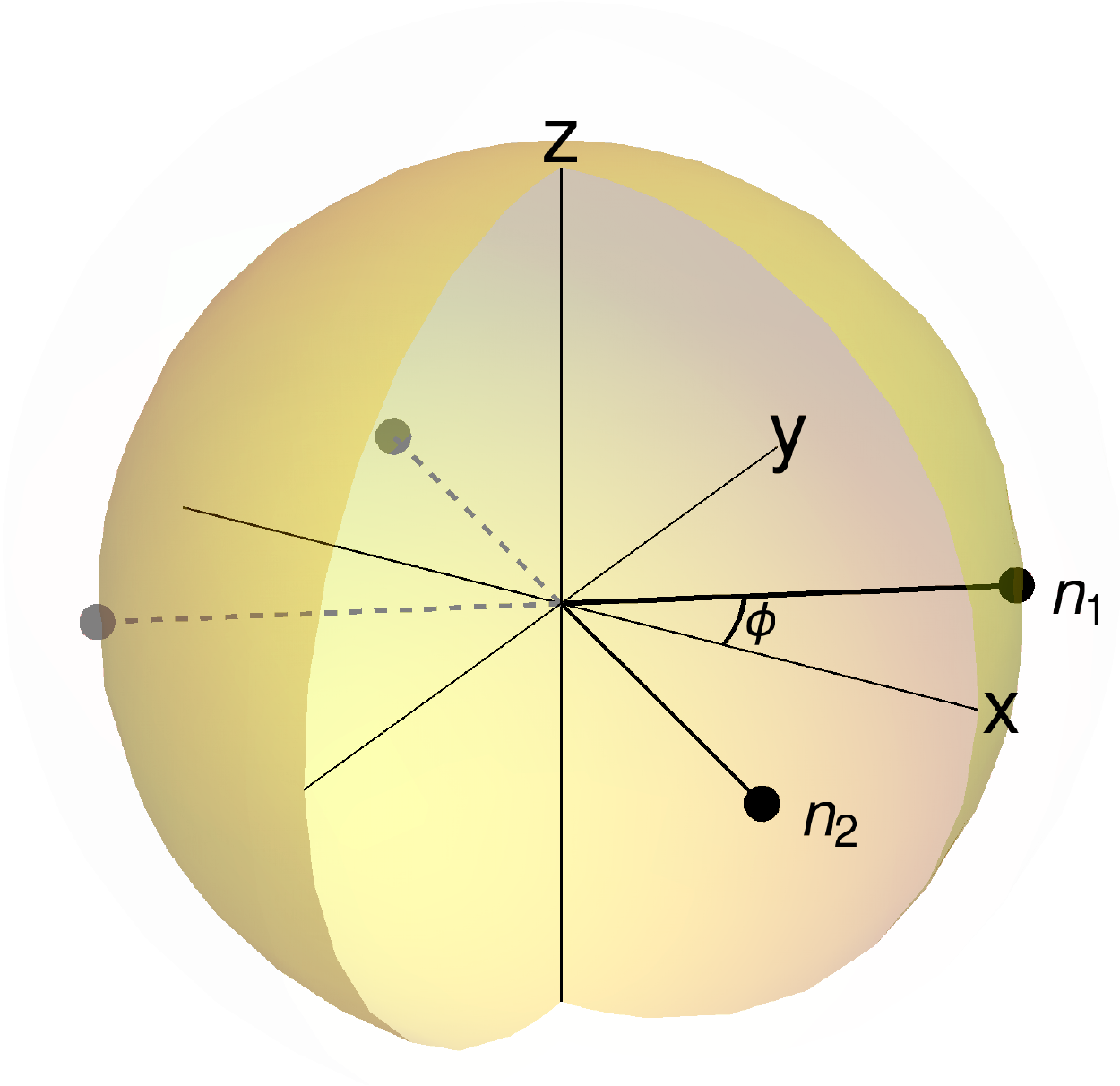}}  
\caption{\label{S1.Rect} The constellation $\co_{\rho}^{(2)}$ of $\rho$ for $\si=2$ and $s=1$.  $\co_{\rho}^{(2)}$ is oriented such that the constellation lies in the $xy$-plane. The black points are an element of the class $[\sco_{\rho}^{(2)}]$. 
}
\end{figure}

At first sight, it seems that how we deal with the relative phase factors in \eqref{Ste.mix} via the
subconstellation classes is rather complicated 
compared to other gauges that one
could use. However, as we already mentioned, the phase factors 
can not be invariant under rotations, and could have complicated transformations
laws in other gauges as well. The main advantage to associate the relative phase factors with
the subconstellation classes is that their transformation laws under rotations
are the same as for all the subconstellations. In addition, when one parametrizes
the whole set of density matrices, the subconstellation classes can be also
counted. 
 Let us discuss this at the hand of the $s=1$ case. Here the states are labeled with two radii
$w_{\si}$ $(\si=1,\,2)$ and they have two associated constellations: $\co^{(1)}$ is the
pair of antipodal points $\pm \bm{r}$ with subconstellation classes
$[\bm{r}]$ and $[-\bm{r}]$ and $\co^{ (2)}$ is given by two axes
that span a rectangle (see Fig. \ref{S1.Rect}) with classes $[\bm{n_1} ,
\bm{n_2}] = [-\bm{n_1} , -\bm{n_2}] $ and $[-\bm{n_1} , \bm{n_2}] =
[\bm{n_1} , -\bm{n_2}] $. Let us orient the coordinate system such
that the sides of the rectangle $\co^{(2)}$ are parallel to the
$\bm{x}$ and $\bm{y}$-axes. We denote by $\phi$ the angle between
the $\bm{x}$-axis and the star $\bm{n}_1$ in the first 
quadrant and specify the class $[\sco^{(2)}]$ with the
vectors $\bm{n}_1$ and $\bm{n}_2$ (see Fig. \ref{S1.Rect}). 
As we can observe, to parametrize all the possible classes $[\sco^{(2)}]$ we must consider $\phi \in [0,\pi/2]$. The associated vectors of the subconstellations are 
\begin{align}
\tilde{\bm{\rho}}^{(1)}= & N_1\left( -r_x + i r_y , \sqrt{2} r_z, r_x + i r_y \right) \, ,
\nn
\\
\tilde{\bm{\rho}}^{(2)}= & N_2 \left(1 ,0, -\frac{2}{\sqrt{6}}\cos (2\phi) , 0 , 1 \right)  
\, ,
\end{align}
with 
\begin{equation}
N_1 = \frac{1}{\sqrt{2}r} \, , \quad N_2= \left(
2 + \frac{2}{3} \cos^2 (2 \phi)
 \right)^{-1/2} \, .
\end{equation}
Therefore, we have parametrized the whole set of spin $s=1$ states
modulo the semidefinite positive condition. 

The first question regarding the semidefinite positive condition is
whether there is a set of classes $\{[\sco^{ (\si)}] \}_{\si=1}^{2s}$ such that for any possible radii $w_{\si}$, the respective density matrix does not represent a physical state.  
We can prove that in a ball close enough to the maximally mixed state
$\rho^* = (2s+1)^{-1} \Id$, there exist states with any subconstellation classes $\{[\sco^{ (\si)}] \}_{\si=1}^{2s}$. The statement is proved by Mehta's lemma (\cite{Bengtsson17} p. 466): 
\begin{lemma}
Let $M$ be a Hermitian matrix of size $D$ and let $\de = \Tr \, M / \sqrt{\Tr \,( M^2)}$. If $\de \geq \sqrt{D-1}$ then $M$ is positive.
\end{lemma}

For a density matrix \eqref{Ste.mix}, $\de= \left( \sum_{\si=0}^{2s} w_{\si}^2 \right)^{-1/2}$ and hence if
\begin{equation}
\sum_{\si=1}^{2s} w_{\si}^2 \leq \frac{1}{2s(2s+1)} \, , 
\label{Mehta.ineq}
\end{equation}
then $\rho$ represents a physical state, independent of its subconstellation classes $[\sco^{(\si)}_{\rho}]$. 
\begin{table*}[t!]
\large
\begin{tabular}{c|c|c|c|c|c|c|c|c|c|}
\cline{2-10}
& \multicolumn{3}{c|}{SC }  &  \multicolumn{3}{c|}{GHZ } &  \multicolumn{3}{c|}{W } 
\\
\cline{2-10}
&  $\quad  \,  \, w_1 \, \, \quad $ & $\quad \, \,  w_2 \, \, \quad $ & $\quad \, \, w_3 \, \, \quad $ 
&  $\quad  \,  \, w_1 \, \, \quad $ & $\quad \, \,  w_2 \, \, \quad $ & $\quad \, \, w_3 \, \, \quad $ 
&  $\quad  \,  \, w_1 \, \, \quad $ & $\quad \, \,  w_2 \, \, \quad $ & $\quad \, \, w_3 \, \, \quad $ 
\\
\hline
\multicolumn{1}{|@{}c@{}|}{$\rho$} 
& $\frac{3}{2\sqrt{5}}$ & $\frac{1}{2}$ & $\frac{1}{2\sqrt{5}}$ & $0$ & $\frac{1}{2}$ & $\frac{1}{\sqrt{2}}$& $\frac{1}{2\sqrt{5}}$ & $\frac{1}{2}$ & $\frac{3}{2\sqrt{5}}$
\\[5pt]
\hline
\multicolumn{1}{|@{}c@{}|}{$\rho_{1}$} 
& $\frac{1}{\sqrt{2}}$ & $\frac{1}{\sqrt{6}}$ & $ $ & $0$ & $\frac{1}{\sqrt{6}}$ & $ $ & $\frac{1}{3\sqrt{2}}$ & $\frac{1}{\sqrt{6}}$ & $ $
\\[5pt]
\hline
\multicolumn{1}{|@{}c@{}|}{$\rho_{1/2}$}  
& $\frac{1}{\sqrt{2}}$ & $ $  & $ $ 
& $0$ & $ $  & $ $ 
& $\frac{1}{3\sqrt{2}}$ & $ $  & $ $ 
\\[5pt]
 \hline
 &  \multicolumn{3}{@{}c@{}|}{  \scalebox{0.4}{\includegraphics{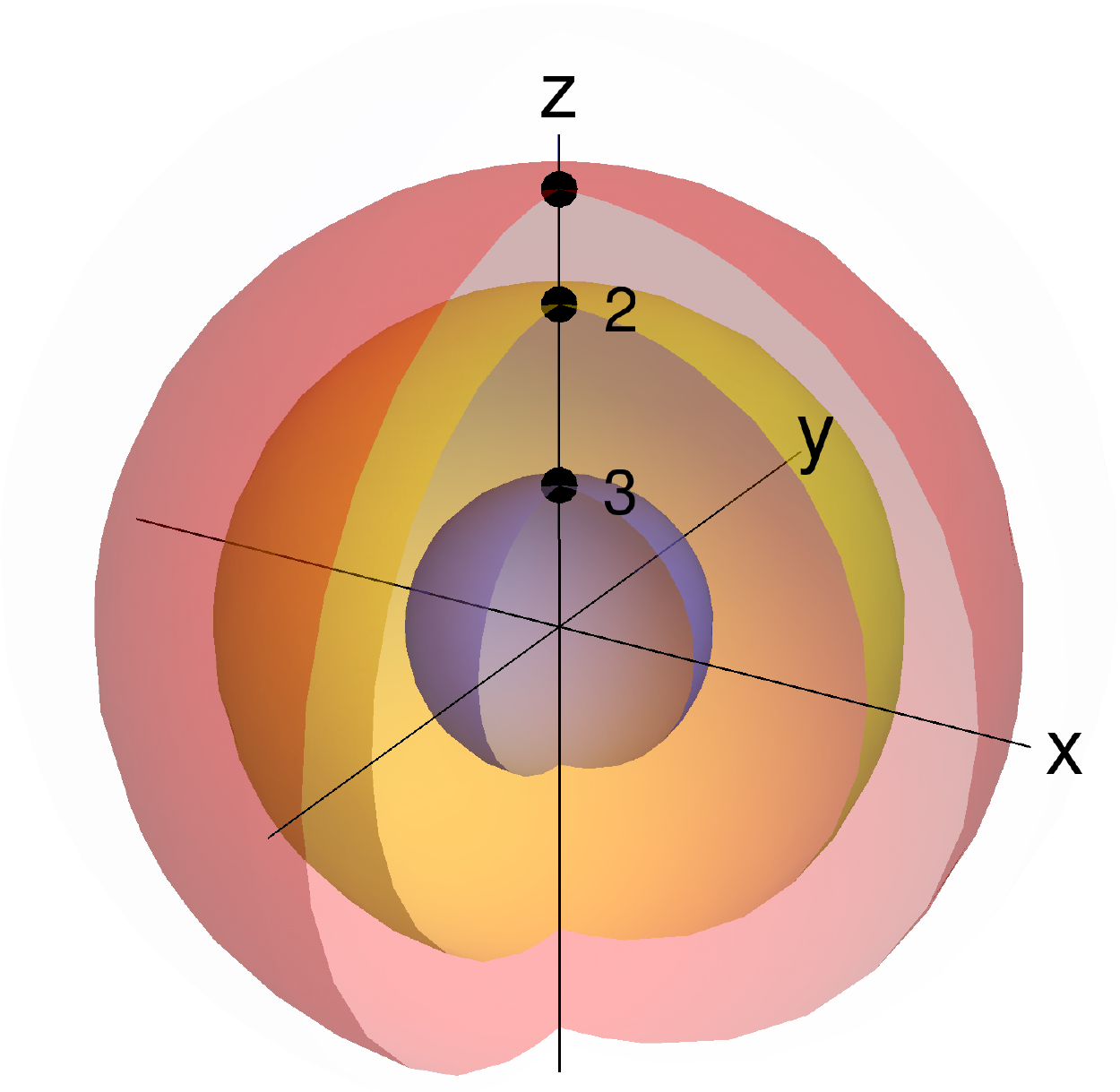}} } 
& \multicolumn{3}{@{}c@{}|}{\scalebox{0.4}{\includegraphics{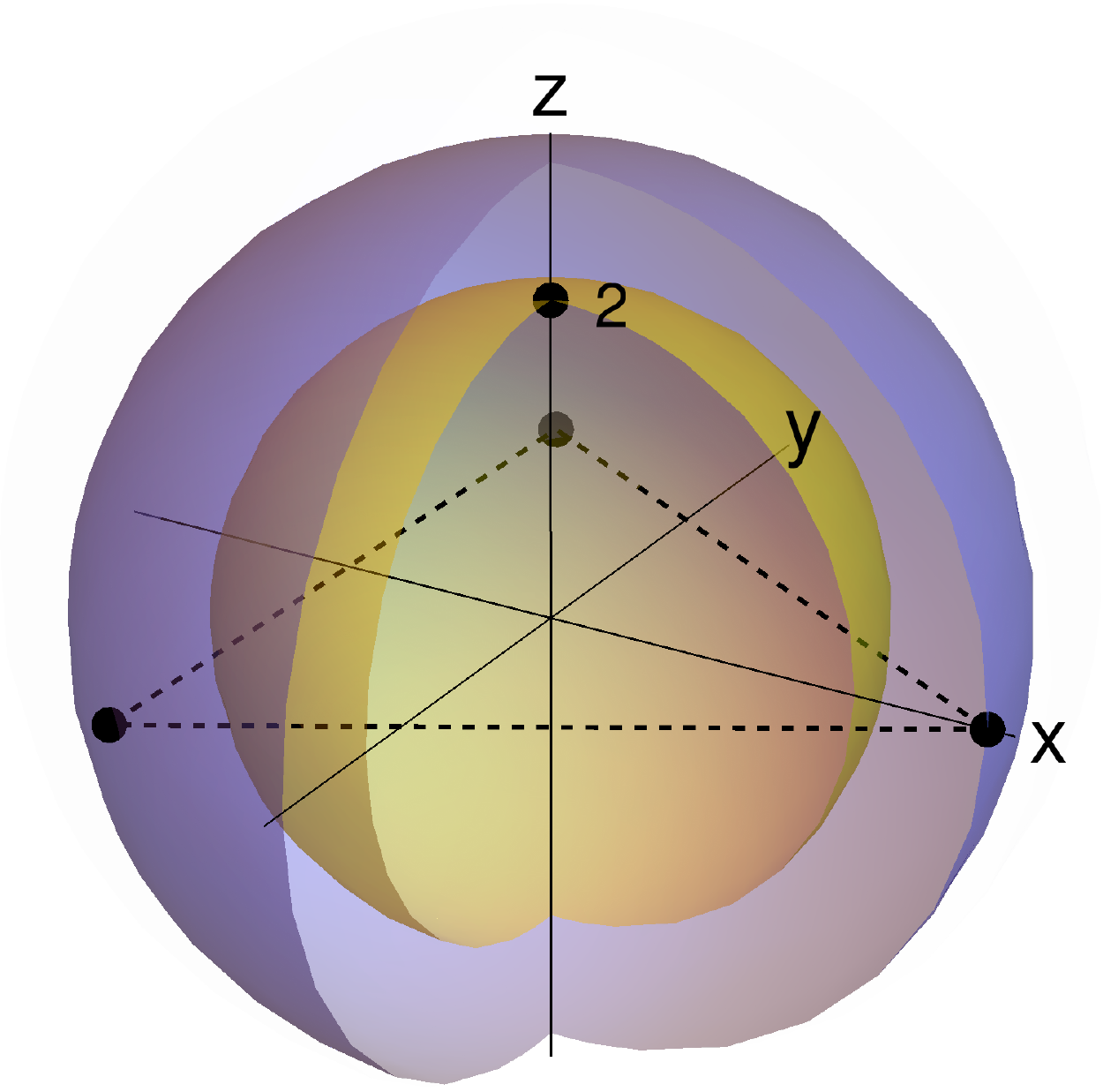}} } 
& \multicolumn{3}{@{}c@{}|}{\scalebox{0.4}{\includegraphics{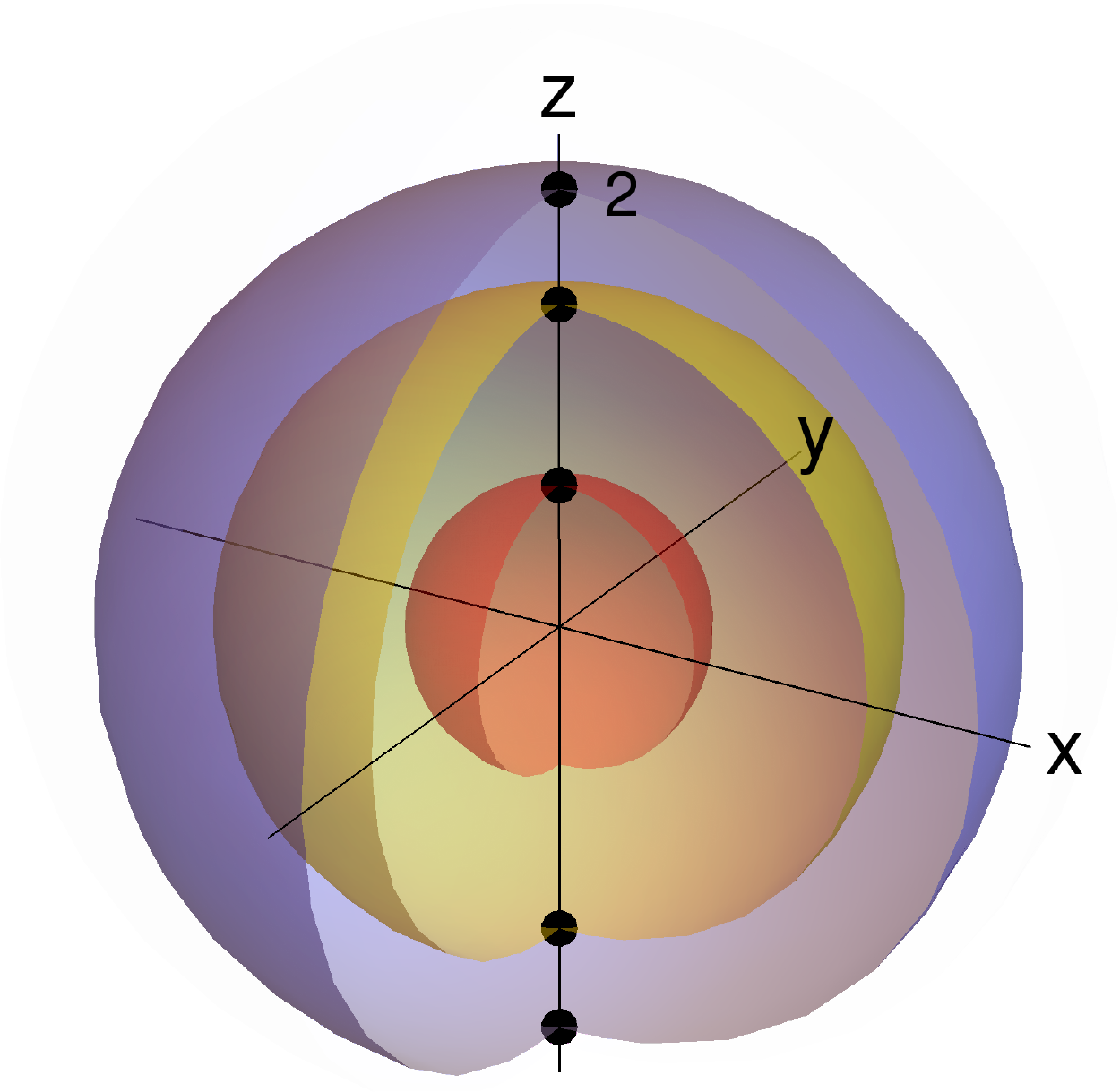}} } 
\\
\cline{2-10}
\end{tabular}
\caption{\label{MSC.rho}
The $T$-representation of the SC, GHZ and W states for
$s=3/2$ in a particular orientation. \emph{Top:} The radii
$ w_{\si}$ of $\rho$ and  their reduced density matrices
$\rho_1$ and $\rho_{1/2}$ after loss of one or two particles (Eq. \eqref{scale.factor}),
respectively. 
\emph{Bottom:} A representative element of each subconstellation class $\{ [\sco_{\rho}^{ (\si)}]\}_{\si}$ for the three states where for each value of $\si=1,2,3$ we assign the color red,
yellow and blue to the respective sphere with radius $w_{\si}(\rho)$. We add the degeneracy number of each star in case it is degenerate. The reduced density matrices $\rho_k$ inherit the
constellations of $\rho$ up to $\si=2k$ with different radii. }
\end{table*}
\subsubsection*{Examples}
Let us study some spin-state families. Some of these families are also described in \cite{Ash.Sir:13} using the $T$-rep without taking into account the subconstellation classes.

\emph{Spin Coherent (SC) states}: Let us consider first the state $\ket{\psi} = \ket{s,s}$, which is the SC state pointing in the $\bm{z}$ direction. We use the expression of \eqref{decomp.TensOp} to obtain the decomposition of $\rho= \ket{s,s}\bra{s,s}$
\begin{equation}
\label{SC.state}
\rho_{ \sigma \mu} = 
\delta_{\mu , 0} (2s)!  \left[ \frac{2\si+1}{(2s+\si+1)!(2s-\si)!} \right]^{1/2} \, .
\end{equation}
Therefore 
\begin{itemize}
\item The components of $\bm{\rho}_{\si}$ are zero except $\rho_{\si 0}$.
\item Every constellation $\co^{(\si)}$ has $\si$ stars in each Pole, which are the simplest constellations with antipodal symmetry.  
\item $[\sco^{ (\si)}]= [\bm{z} , \, \dots , \, \bm{z}]$, \ie, an
    element of the class $[\bm{c}^{(\si)}]$ is the subconstellation formed by $\si$ stars along the $\bm{z}$ direction. 
\item The radii $w_{\si}$ have the values
\begin{equation}
w_{\si}= (2s)!  \left[ \frac{2\si+1}{(2s+\si+1)!(2s-\si)!} \right]^{1/2} \, .
\end{equation}
\end{itemize}
The density matrix of the pure SC state in direction $\bm{n}(\theta, \phi)$ is obtained rotating the state $\rho_{\bm{z}}$ by a rotation with Euler angles $(\phi, \, \theta , \, 0)$. Using the equations in \cite{Var.Mos.Khe:88} ( p. 113), we obtain that
\begin{align}
\rho_{\bm{n}} =& \rho_{00} T_{00} + \sum_{\sigma=1}^{2s} w_{\si} \sum_{\mu=-\si}^{\si} D_{\mu , 0}^{(\si)}(\phi,\theta,0) T_{\sigma \mu} \, ,
\\
=& \frac{\mathds{1}}{2s+1} + \sum_{\sigma=1}^{2s} w_{\si} \sqrt{\frac{4\pi}{2\si+1}} \sum_{\mu=-\si}^{\si} Y^*_{\sigma \mu} (\theta , \phi) T_{\sigma \mu} \, ,
\nn
\end{align}
with $Y_{\si \mu}(\theta, \phi)$ the spherical harmonics. The respective subconstellation classes are $[\bm{n}, \, \dots , \, \bm{n}]$ for each $\si$. The states $\ket{\pm \bm{n}} $ only differ by the classes $[\sco^{(\si)}]$ of $\si$ odd (see Eq. \eqref{rel.anti.states}).

\emph{General pure state}: Let us take a spin-$s$ state $\ket{\psi}$ and its density matrix $\rho_{\psi}=\ket{\psi}\bra{\psi}$. The state $\rho$ expanded in the $T$-rep is given by
\begin{equation}
 \rho_{\psi} = \sum_{\si \mu} \bra{\si,\mu} \CP_{\si}  \ket{ \psi,\psi^A} T^{(s)}_{\si \mu} \, ,
\label{pure.state}
\end{equation}
where we use the bipartite notation $\ket{\psi , \, \psi^A} = \ket{\psi} \otimes \ket{ \psi^A}  $ and the antipodal state $\ket{\psi^A}$ is defined as \eqref{antipodal.def}.
We can observe that the constellations of the $T$-rep come from the
irrep decompositions of the bipartite state $ \ket{\psi , \, \psi^A} $, where the antipodal state $\ket{\psi^A}$ appears from the fact that it transforms in
the same way as the bra $\bra{\psi}$ under rotations
\cite{BrinkSatchler68}. In particular, the standard Majorana
constellation $\co_{\psi}$ of the pure spin-$s$ state $\ket{\psi}$ is an element of the class
 $[ \sco^{(2s)} ]$. However, only with the knowledge of the class $[\sco^{(2s)}]$, we cannot specify the state $\ket{\psi}$. An algorithm to recover the standard Majorana polynomial from $[\sco_{\rho_{\psi}}^{(2s)}]$ is the following: calculate the overlap between $\rho_{\psi}$ and the SC states pointing 
to a star $\bm{n}$ of an element of $[\sco^{ (2s)}]$. If $\braket{\bm{n}|\rho_{\psi}|\bm{n}}=0$, then $-\bm{n} \in \co_{\psi}$,
otherwise $\bm{n} \in \co_{\psi}$.  

\emph{Dicke state}: The Dicke states $\rho_m=\ket{s,m}\bra{s,m}$ with
$m=-s,\dots ,s$ satisfy {$\Tr (\rho_m T_{\si \mu} )=0$} for
$\mu\neq 0$. 
For $\rho_m$, $ w_{\si} = |\braket{s , \, m | T_{\si 0} |s , \, m }| =
|C^{\si 0}_{s m\, s -m}|$. 
We conclude the following results:
\begin{itemize}
\item The constellations $\co_{\rho_m}^{ (\si)}$ are the same for all $m=-s,\dots s$, with $\si$ stars in the each Pole.
\item The respective classes $[\sco^{ (\si)}]$ are obtained calculating the sign of the coefficients in \eqref{pure.state},
\begin{align}
\bra{\si,\mu } P_{\si} \ket{ \psi,\psi^A} = (-1)^{s-m} \delta_{\mu 0} C^{\si 0}_{s m s -m} \, .
\end{align}
In Table \ref{MSC.rho} we observe the Dicke states for spin-$3/2$ states which are equivalent up to a rotation to the SC and W states. 
\item  The antipodal states $\rho_m$ and $\rho_{-m}$ just differ by some classes $[\bm{c}^{(\si)}]$ of $\si$ odd, as we show in \eqref{rel.anti.states}.
\end{itemize}
\subsection{The polynomials of $T_{\si \mu}$}
The polynomials of the tensor operators $T_{\si \mu}^{(s)}$ are the
irreps of $SU(2)$ in $P^{(N,N)}(z)$, where one can compare and  
multiply polynomials of different degrees (\ie, elements of different spaces
$P^{(N,N)}(z)$) more easily than for their matrix counterpart, which involves tensor product and projections in the fully symmetric sector. The first result regarding this property is associated with the comparison of the tensor operators of different spin-$s$. Before explaining the general results, let us compare the Majorana polynomials for $T^{(s)}_{1 0}$ for $s=1/2,1$. From equation \eqref{ms.pol} we obtain that
\begin{equation}
p^{(1)}_{10} (z) = \left( z^a z_a \right) p^{(1/2)}_{10}(z) \, .
\end{equation}
We observe that the binomial $(z^a z_a)$ is the factor between
polynomials representing the same operator but for  different spin. In
addition, it is easy to observe that $L (p_{10}^{(1/2)})=0$. We
summarize these results in the following theorem. Its proof can be
found in Appendix \ref{Proof.BigT}. 
\begin{theorem}
\label{Big.Theo}
The polynomials  $p_{\si\mu}^{(s)}(z)$ associated to the operators $T^{(s)}_{\si \mu}$ have the following properties
\begin{enumerate}
\item The action of the partial trace operator $L$ under $p_{\si \mu}^{(s)}(z)$ is equal to 
\begin{equation}
L \left( p_{\si \mu}^{(s)}(z) \right) = \left\{ 
\begin{array}{c c}
\frac{\sqrt{(2s+\si+1)(2s-\si)}}{2s} p_{\si \mu}^{(s-1/2)}(z) & \text{ if } s >\si/2
\\
0 & $otherwise$
\end{array}
 \right.  . 
\label{factor.L}
\end{equation}
In particular,
\begin{equation}
(2s+1)^{-1/2} L\left(p_{00}^{(s)}(z)\right) = (2s)^{-1/2} p_{00}^{(s-1/2)}(z) \, ,
\end{equation}
and therefore $L$ leaves the trace of the respective operator invariant.
\item For any value of $\si \leq 2s$, 
\begin{equation}
p_{\si \mu}^{(s)} (z) = l(s,\si)^{-1} (z^a z_a)^{2s-\si} p_{\si \mu}^{(\si/2)}(z)  \, ,
\end{equation}
with
\begin{align}
l(s,\si) \equiv & \sqrt{\frac{(2s+\si+1)!(2s-\si)!}{(2\si+1)!}} \frac{\si !}{(2s)!} \, .
\end{align}
\end{enumerate}
\end{theorem}
\begin{figure*}[t!]
\large
\begin{tabular}{|c|c|c|}
\hline
$s=5/2$  & $s=3$ $\rho^C$ &  $s=3$ $\rho^Q$ 
\\
\hline
 \scalebox{0.43}{\includegraphics{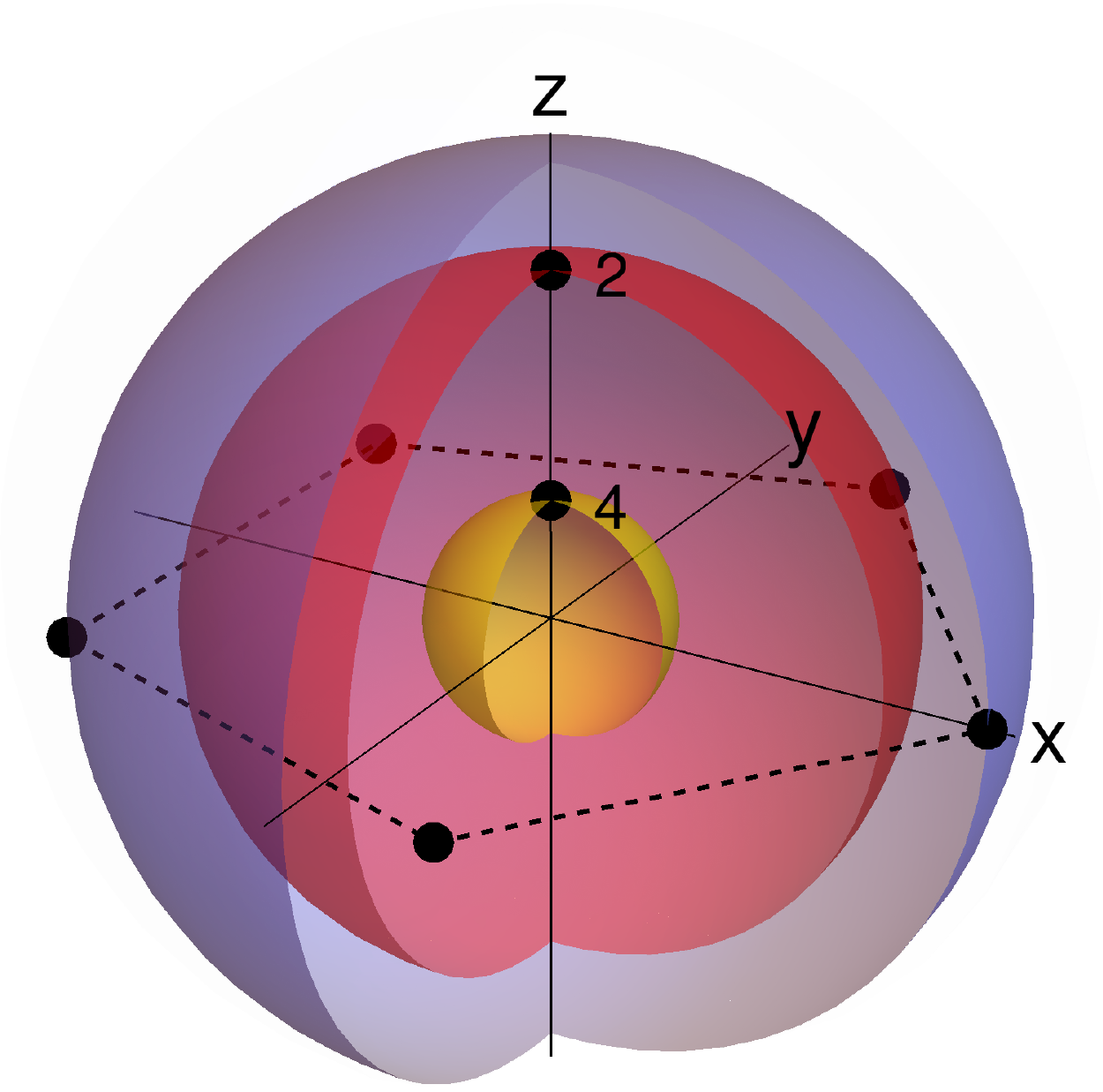}}  
& \scalebox{0.43}{\includegraphics{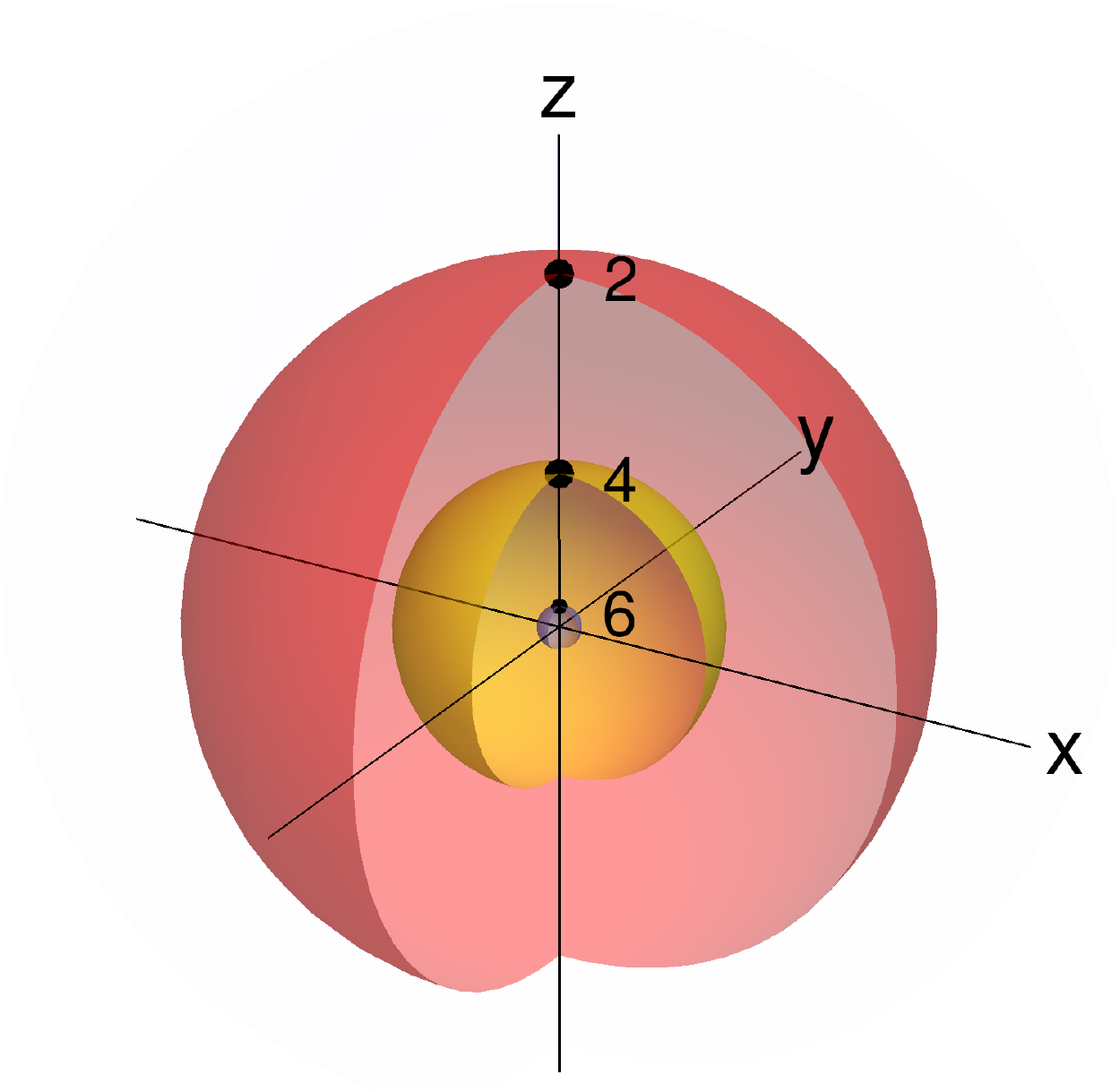}}  
& \scalebox{0.43}{\includegraphics{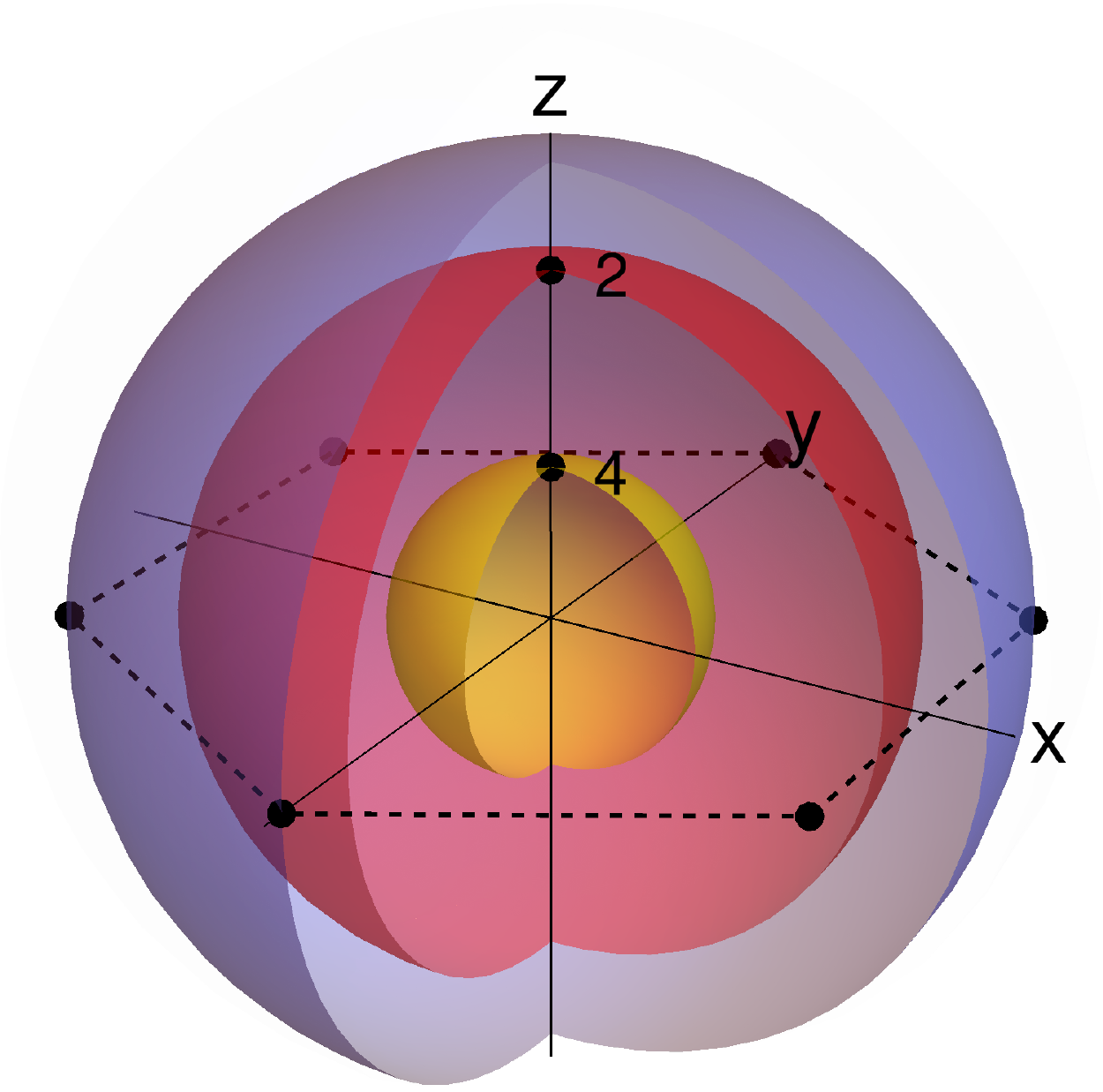}} 
\\
\hline
\end{tabular}
\caption{\label{MSC.SCS}
$T$-rep for the density matrices of Schr\"{o}dinger cat states $\rho^Q$ and classical cat states $\rho^C$ for $s=5/2,3$. \emph{Left:} $\rho^Q$ for $s=5/2$. $\rho^C$ is equal to
$\rho^Q$  with the highest class $[\sco_{\rho^Q}^{(2s)}]$ taken out, which corresponds in the figures to the stars lying on the blue sphere. 
\emph{Center:} $\rho^C$ for $s=3$. \emph{Right:} 
$\rho^Q$ for $s=3$. The states $\rho^C$ and $\rho^Q$ differ only by $[\sco^{(2s)}]$: For integer spin $s$, the highest constellation for $\rho^C$ shrinks to a small value given by eq.\eqref{wnC}, whereas for half-integer $s$ the radius of the sub-constellation vanishes and hence does not contribute.  For any spin value, the states $\rho^C$ and $\rho^Q$ after the partial trace of one of its constituent spins-1/2 are the same, $\rho^Q_{s-1/2}=\rho^C_{s-1/2}$.} 
\end{figure*}
\subsubsection*{Inherited constellations in the $T$-rep}
\label{sec.tensorop}
 By construction, the action of a $SU(2)$ transformation on $\rho$ rigidly rotates all its classes $[\sco_{\rho}^{(\si)}]$ while their radii $w_\sigma$ are invariant. In addition to the well-behaviour under rotations of the $T$-rep and a visual representation of our states (see Table \ref{MSC.rho}), there is an additional property associated to their reduced matrices $\rho_k$. From  Theorem \ref{Big.Theo}, the spin-$s'$ (with $s'=s-1/2$) reduced state $\rho_{s'} = \Tr_1 (\rho )$ is equal to
\begin{equation}
\label{red.state}
\rho_{s'} = \sum_{\si=0}^{2s-1} \frac{\sqrt{(2s+\si+1)(2s-\si)}}{2s} w_{\si} \tilde{\bm{\rho}}_{\si } \cdot \bm{T}_{\si}^{(s')} \, .
\end{equation}
As we can observe, each component is re-scaled by a factor independent of $\mu$ leaving the subconstellation classes invariant, \ie, the reduced density matrices inherit the lowest classes of $\rho$, $\{  [\sco^{(\si)}_{\rho_{s'}} ]\}_{\si=1}^{s'}=\{ [\sco^{(\si)}_{\rho}]\}_{\si=1}^{s'}$. The re-scaled factor can be absorbed in the radius $w_{\si}$
\begin{equation}
w_{ \si}(\rho_{s'} ) =  \frac{ \sqrt{(2s+\si+1)(2s-\si)} }{2s}  w_{\si}(\rho) \, ,
\label{scale.factor}
\end{equation}
where we write the weights as a function of the density matrix. The radius $w_{\si}$ increases with respect to a particle loss if $ \si (\si+1) < 2s $. If the state loses more than one particle, the lowest classes are still inherited and the radii are re-scaled with a
product of factors of the front of \eqref{scale.factor} with successively reduced spin $s$. In Table \ref{MSC.rho} we plot the radii and classes of $\rho$ $\{w_{\si} , [\sco^{ (\si)} ]\}$ for $\rho$ equal to the SC, GHZ and W states with $s=3/2$. We only plot a representative element of $[\sco^{ (\si)}]$ to simplify the visualization in the figures. The table also includes the radii $w_{\si}$ for each reduced density matrix $\rho_{k}$ with $k=1/2,1$. 

To study another example, let us discuss the constellation
differences between the quantum linear superposition $\rho^Q=
\ket{\psi} \bra{\psi}$ with $\ket{\psi}= (\ket{s,s} +
\ket{s,-s})/\sqrt{2}$ (a ``Schr\"{o}dinger cat'' state) and a classical
mixture of the same states $\rho^C = (\ket{s,s} \bra{s,s} + \ket{s,-s}
\bra{s,-s})/2$ (which we will call ``classical cat state'' for short). $\rho^{Q}$ has an
additional term with respect to $\rho^C$,  
\begin{align}
\rho^{Q} = & \rho^C + \frac{1}{2}(\ket{s,s}\bra{s,-s} + \ket{s,-s}\bra{s,s})
\nn
\\
= &  \rho^C + \frac{1}{2}( (-1)^{2s} T_{2s,2s}^{(s)} + T_{2s,-2s}^{(s)}) \, ,
\label{QvsC.cat}
\end{align}
and it yields that the constellations set of these two states will be equal except for $\co^{(N)}_{\rho}$, and hence $[\sco_{\rho}^{(N)}]$, with $N=2s$. Let us calculate
the constellations of $\rho^C$ first. Using equation \eqref{SC.state} 
 and that $ \ket{s,-s} \bra{s,-s}= A \ket{s,s} \bra{s,s} A^{\dagger}$, we obtain that 
\begin{equation}
(\rho^C )_{\si \mu} =  \left\{
\begin{array}{c c}
\delta_{\mu,0} (2s)!\left[ \frac{2\si+1}{(2s+\si+1)!(2s-\si)!} \right]^{1/2}  & \si \, \, \text{even} 
\\
0 & \si \, \, \text{odd}
\end{array} \right. \, ,
\end{equation}
and hence $\rho^C$ does not have constellations for $\si$ odd. In particular, $\co^{(N)}_{\rho^C}$ for $N$ odd does not exist and for $N$ even it is equal to $N$ points in each Pole. On the other hand, the vector $\bm{\rho}^{(N)}$ of $\rho^Q$ and the respective polynomials are given by
\begin{equation*}
\bm{\rho}^{(N)}_{\rho^Q} = \left\{
\begin{array}{c c}
\left( \frac{1}{2} , 0, \dots ,0 , \frac{(2s)!}{\sqrt{(4s)!}} ,0, \dots,0 , \frac{1}{2}  \right) &  \text{for} \quad N \, \text{even} 
 \\
\left( \frac{(-1)^{2s}}{2} , 0, \dots ,0 , \frac{1}{2}  \right) & \text{for} \quad N \, \text{odd} 
\end{array}
\right. ,
\end{equation*}
\begin{equation}
p^{(N)}_{\rho^Q}(\zeta) = \left\{
\begin{array}{c c}
\frac{1}{2} ( z^{2s} + 1)^2    &  \text{for} \quad N \, \text{even} 
 \\
\frac{(-1)^{2s}}{2} (z^{4s} + 1)  & \text{for} \quad N \, \text{odd} 
\end{array}
\right. \, .
\label{cons.Qcase}
\end{equation}
The roots of the polynomials \eqref{cons.Qcase} draw  on the sphere a $4s$-agon in the odd case and a $2s$-agon with all the stars doubly-degenerate in the even case. The radii $ w_{N}$ for each case are equal to 
\begin{align}\label{wnC}
w^C_{N} = & \left\{ 
\begin{array}{c c}
 \frac{(2s)!}{\sqrt{(4s)!}}   &  \text{for} \quad N \, \text{even} 
 \\
0  & \text{for} \quad N \, \text{odd} 
\end{array} \right. \, ,
\nn
\\
 w^Q_{N} =& \left\{ 
\begin{array}{c c}
 \sqrt{ \frac{1}{2} + \frac{((2s)!)^2}{(4s)!} }   &  \text{for} \quad N \, \text{even} 
 \\
\frac{1}{\sqrt{2}}  & \text{for} \quad N \, \text{odd} 
\end{array} \right. \, .
\end{align} 
Our calculations are in agreement with the results in
  \cite{Ash.Sir:13} were the authors also calculated  the constellations of the classical and quantum cat states for a general spin value $s$. In figure \ref{MSC.SCS} we plot the states $\rho^Q$ and $\rho^C$ for $s=5/2,3$ with an element of their respective classes $[\sco^{(\si)}]$. In addition, by the results of the previous subsection, the states after the reduction of one constituent spin-$1/2$ have the same subconstellation classes and radii and therefore they are equal, $\rho_{s-1/2}^Q=\rho_{s-1/2}^S$. As a consequence, we obtain the old known result that the GHZ state after the loss of a particle is separable \cite{Duer2000}.
\subsection{Tensor product and the $S$-rep}
 Some operators $C \in \Bs (\Hs_s)$ are the projection of the tensor product of $N$ spin-$1/2$ operators $C=\CP_s C_1 \otimes \dots \otimes C_N \CP_s$, where, again, the projector operator $\CP_s$ is considered to be restricted to its image. The polynomials of these operators are factorizable
 \begin{equation}
 p_C(z) = \prod_{k=1}^{N} p_{C_k}(z) \, ,
\label{fac.pol}
 \end{equation}
where the proof consists in the calculation of $\braket{-\bm{n}_B |
  \CP_s C \CP_s | -\bm{n}_B}$ in terms of the symmetric Dicke states
\eqref{Dicke.state}. In particular, the set of operators given by the
tensor product of $N$ Pauli matrices $\si_{\mu}$ with $\mu=0,x,y,z$
projected in the fully symmetric space is a tight frame of $\Bs
(\Hs_s)$ \cite{Gir.Bra.Bag.Bas.Mar:15} that we called the
\emph{S-rep}. In an equivalent way and following the same reasoning as
in \cite{Gir.Bra.Bag.Bas.Mar:15}, the set of projected tensor products
of the spin-$1/2$ operators $\{\si_0 , \, \si_-, \, \si_z , \, \si_+
\}$, $S_{\tau_1 \dots \tau_N} \equiv  \CP_s \si_{\tau_1} \otimes \dots
\otimes \si_{\tau_N} \CP_s$ with $\tau_k=0, \, - , \, z , \,+$, is a
tight frame. The operator $S_{\tau_1 \dots \tau_N}$ is independent of
the order of its indices $\tau_k$, and the only relevant information
can be encoded in a 4-vector of natural numbers $\vec{\nu}=(\nu_0 ,
\, \nu_{-} , \, \nu_z , \, \nu_{+})$, where $\sum_j \nu_j = N$ and
$\nu_j$ is the number of times that $j$ appears in the indices of
$S_{\tau_1 \dots \tau_N} $. Following the previous result, the
polynomial of $S_{\vec{\nu}}$ is factorized in powers of the
polynomials of $\si_j$ with $j=0,\, \pm , \, z$,  
\begin{equation}
p_{S_{\vec{\nu}}} (z) = \prod_{j} (p_{j}(z))^{\nu_j} \, .
\label{TP.pol}
\end{equation}
\subsection{Connection between $T$- and $S$-reps}
 In this subsection we will obtain an explicit formula for writing
 the $T_{\si \mu}$ operators in terms of the $S$-rep, using their respective polynomials. The operators in the $T$-rep and $S$-rep share the property that their polynomials contain the factor $(z_a z^a)^{k}$, where $k=2s- \si$ for $T^{(s)}_{\si \mu}$ and $k=\nu_0$ for $S_{\vec{\nu}}$. Both of them are a basis of $\Bs (\Hs_s)$. In particular, the operators $T_{\si \mu}^{(s)}$ can be written in terms of the $S$-rep 
\begin{equation}
T_{\si \mu}^{(s)} = \sum_{\vec{\nu}} A_{\si \mu}^{\vec{\nu}} S_{\vec{\nu}} \, .
\end{equation}
Lemma \ref{prod.op} and Theorem \ref{Big.Theo} yields that
\begin{align}
\Tr \left( \left( T^{(s)}_{\si \mu} \right)^{\da} S_{\vec{\nu}} \right) \propto & 
\prod_j \left( p_{j}(\partial^b , \pd_b) \right)^{\nu_j} p_{\si \mu}^{(s)}(z_a , z^a)
\label{reduced.spinT}
\\
\propto &
\prod_{j \neq 0} \left( p_{j}(\partial^b , \pd_b)  \right)^{\nu_j} p_{\si \mu}^{(s-\nu_0/2)}(z_a , z^a)
\nn\\
\propto &
 \Tr \left( \left( T^{(s-\nu_0/2)}_{\si \mu} \right)^{\da} S_{(0  ,  \nu_- ,  \nu_z ,  \nu_+)}  \right) \, ,
\nn
\end{align}
and hence $A_{\si \mu}^{\vec{\nu}}=0$ for $\nu_0 > 2s-\si$. The resolution of the $T_{\si \mu}$ operators in the $S$-rep is not unique because the $S$ matrices form a tight frame instead of a basis. However, it is possible to write a resolution only with one running index and $\nu_0 = 2s-\si$ fixed,
\begin{equation}
T_{\si \mu}^{(s)} = 
 \sum_{k = \mu}^{\si} 
 A(s, \, \si , \, \mu , \, k)  S_{(2s-\si, k -\mu, \si + \mu - 2k , k)}  \, ,
\label{exp.TSconnection}
\end{equation}
{with
\begin{align}
 A(s, \, \si , \, \mu , \, k) = 
 \sqrt{
\frac{(\si+\mu)!(\si-\mu)!}{(2\si)!} 
}
\frac{l(s , \, \si)^{-1}(-1)^{k} 2^{\mu-2k}(\si!)}{ k! (k-\mu)! (\si+\mu-2k)!} \, .
\label{factor.TSconnection}
\end{align}}
The proof of this equation is in Appendix \ref{App.TScon}. The $S$-rep has also an additional
property under partial traces \cite{Gir.Bra.Bag.Bas.Mar:15}: the
coefficients  
$c'_{(\nu_0 , \, \nu_- , \, \nu_z , \, \nu_+)} = \Tr (\rho_k S_{(\nu_0 , \, \nu_- , \, \nu_z , \, \nu_+)})$ with $\sum_j \nu_j = 2k $ of the  reduced spin-$k$ state $\rho_k$ are equal to a subset of coefficients $c_{\vec{\nu}}$ of the original state $\rho$
\begin{equation}
c'_{(\nu_0 , \, \nu_- , \, \nu_z , \, \nu_+)} = 
c_{(\nu_0  + 2(s-k), \, \nu_- , \, \nu_z , \, \nu_+)} \, .
\label{red.Srep}
\end{equation}
We can prove that the latter result of the $S$-rep is related to the property of the inherited constellations of the $T$-rep discussed in subsection \ref{sec.tensorop} by using the connection between the representations: A state $\rho = \sum \rho_{\si \mu} T_{\si \mu}^{(s)}$ has reduced state $\rho_{s-1/2}$ equal to eq. \eqref{red.state}, and the same eq. \eqref{red.Srep} 
for $k=s-1/2$ can be obtained using that 
\begin{align}
&\Tr \left(T_{\si \mu}^{(s)} S_{(\nu_0 +1 , \, \nu_- , \, \nu_z , \, \nu_+)} \right)
\\
=& \frac{\sqrt{(2s+\si+1)(2s-\si)}}{2s} \Tr \left(T_{\si \mu}^{(s-1/2)} S_{(\nu_0 , \, \nu_- , \, \nu_z , \, \nu_+)} \right) \, .
\nn
\end{align}
The last equation is proved by Theorem \ref{Big.Theo}.
\subsection{Anticoherence order in terms of polynomials}
We end this section writing the criterion for the anticoherent states
in terms of polynomials.  Zimba \cite{Zimba06} defined an anticoherent
state of order-$t$, {or $t$-anticoherent for short}, if the
expectation value $\braket{(\bm{n}\cdot \bm{S})^k}$ is independent of
the unit vector $\bm{n}$ for any $k$ with $0\leq k \leq t$. The
criterion of anticoherence in terms of the $S$- and $T$-
representations were obtained in \cite{Gir.Bra.Bag.Bas.Mar:15}. A
  state is $t$-anticoherent if and only if its spin-$t/2$ reduced
  state $\rho_{t/2}$ is the maximally mixed state $\rho^* =
  (2t+1)^{-1} \Id$ which is equivalent to that $\braket{T_{\si
      \mu}}=0$ for all $1\leq \si \leq t$ and $-\si \leq \mu \leq
  \si$. In terms of the Majorana polynomial of $\rho$, $p_{\rho}(z)$,
  a state $\rho$ is $t-$anticoherent if and only if
  $L^{2s-t}(p_{\rho}(z)) \propto p_{\Id}(z) = (z^a z_a)^{t}$.  
\section{The Husimi- and P-functions of $\rho$}
\label{Sec.IV}
Several quasiprobability distributions are expressed in terms of the coefficients $\rho_{\si \mu}$ \cite{Aga:13}, and we are going to study two of them: The Husimi- and the P-functions \cite{Aga:13}. The Husimi function of a state $\rho$, $H_{\rho}(\bm{n}) \equiv \braket{\bm{n} |\rho | \bm{n}} $, is related to the Majorana polynomial of $p_{\rho}(z)$ as
\begin{equation}
H_{\rho}(-\bm{n})   = \frac{p_{\rho}(z)}{(z^a z_a)^N }  \, ,
\end{equation}
with $\bm{n}$ the direction associated to the complex number $\zeta = z_1 /z_2$ via the stereographic projection. As we can observe, the variables $(z_1 , \, z_2)$ (and hence $(z^a) = (z_a^*)$) are defined up to a common factor. In particular, if one takes $z_1 = \cos(\theta/2)$ and $z_2 = \sin(\theta/2) e^{i\phi}$, the denominator of the last equation is one and hence $H_{\rho}(-\bm{n}) = p_{\rho}(z)$.
On the other hand, the P-function of a state $\rho$ is defined as the
function $P_{\rho}(\bm{n})$ such that
\begin{equation}
\rho = \int P_{\rho}(\bm{n}) \ket{\bm{n}}\bra{\bm{n}} \diff \Omega \, ,
\label{rhoP}
\end{equation}
with $ \diff \Omega$ the volume element of the 2-sphere. The P-function of a state is not unique and the notion of classical
states for spin systems can be expressed in terms of the P-function
\cite{Giraud08}: A state $\rho$ is classical iff a
representation of the form \eqref{rhoP} with non-negative P-function
exists. If one restricts the P-function to a linear combination of the
first $2s$ spherical harmonics $\{Y_{\si \mu} (\theta ,
\phi)\}_{\si=1}^{2s}$, one obtains a unique P-function for each state
\cite{Aga:13}  
\begin{equation}
P_{\rho}(\theta , \, \phi) \equiv \sum_{\si=0}^{2s} \sum_{\mu} \frac{(-1)^{\si-\mu} l(s,\si) \sqrt{(2\si+1)!}}{\sqrt{4\pi} (\si!)} \rho_{\si \mu} Y_{\si \mu} (\theta, \phi)  \, .
\end{equation}
Using Theorem \ref{Big.Theo}, we can calculate the $P$-function of the spin-$k$ reduced density matrices $\rho_k$ in terms of the coefficients of the original state $\rho$, yielding that
\begin{equation}
P_{\rho_k}(\theta , \, \phi) = \sum_{\si=0}^{2k} \sum_{\mu}  \frac{(-1)^{\si-\mu} l(s,\si) \sqrt{(2\si+1)!}}{\sqrt{4\pi} (\si!)} \rho_{\si \mu} Y_{\si \mu} (\theta, \phi) \, ,
\end{equation}
\ie, the P-function of the reduced density matrices is equal to the P-function of the original state omitting the higher multipolar terms.
\section{Summary and concluding remarks}
\label{Sec.Con}
We have generalized the Majorana stellar representation of pure states
to Hermitian operators, in particular to 
density operators and hence mixed states. The mapping is a bijective
correspondence between states $\rho \in \Bs (\Hs_s)$, polynomials
$p_{\rho}(z) \in P^{(N,N)}(z)$ and a set of subconstellation classes
on the Euclidean space $\mathds{R}^3$, where the latter is equal to
the Ramachandran-Ravishankar representation  
\cite{Ram.Rav:86}, called here the $T$-rep. The representation behaves
well under rotations by construction. In addition, it has also attractive properties such as: definition of polynomials for any operator $C\in \Bs (\Hs_s)$; inherited constellations under partial traces; the tensor product of operators in the fully symmetric sector is reduced to the
product of their polynomials;  and any other operation in $\Bs
(\Hs_s)$ can be written as differential operation acting on the corresponding polynomials.
Some of these results have been found previously in the $T$- and $S$- representations, and now, with the Majorana polynomial, the bridge between them has been explained and their results can be translated from one to another. In addition, we discussed the $T$-representation in terms of subconstellation classes that allows us to completely follow the state under rotations and, with the results derived here, also under the partial trace. Each subconstellation class represents the $\si$-block of the state $\rho$, and its radius  $w_\si$ represents its magnitude. The states written in the $T$-rep have been used to study the quantum polarization of light \cite{Hoz.Kli.Bjo:13}. The results presented here helps to represent each block easily and track its changes under partial traces. We also wrote the relation between the Majorana representation of a state $\rho$ and its Husimi and P-functions. We hope that this new representation, as the standard Majorana representation for pure states, can give the readers more intuition about the space of the mixed states and the action of the $SU(2)$ group on it.  
\section*{Acknowledgements}
ESE thanks the University T\"ubingen and its T@T fellowship. The authors thank John Martin for fruitful correspondence.
\begin{appendix}
\section{Proofs of some Lemmas}
\label{proof.lemmas}
\emph{Proof of Lemma \ref{Action.optoop}}. Let us consider first the polynomials
$p_D(z')$ and $p_E(z)$ written in different variables, with product equal to
\begin{equation}
p_D(z') p_E(z) = \braket{-\bm{n}_B' | D | -\bm{n}_B'}  \braket{-\bm{n}_B | E | -\bm{n}_B} \, .
\label{two.zetas}
\end{equation}
To obtain the polynomial of $C=DE$, $p_C(z)$, we have to apply a differential operator $\Op$ dependent only 
on the variables $z^{a'}$ and $z_{a}$, such that 
$\Op( \ket{{-\bm{n}_B'}}  \bra{-\bm{n}_B}   ) = \Id$. Note that $\ket{{-\bm{n}_B'}}  \bra{-\bm{n}_B}$
can be seen as a matrix with entries 
\begin{align}
&\braket{ s , m' |-\bm{n}_B'}  \braket{-\bm{n}_B | s, m} = (-1)^{2s-m-m'} 
\\
& \times \sqrt{\binom{2s}{s-m} \binom{2s}{s-m'}}
z_1^{s+m} z_2^{s-m} (z^{1\prime})^{s+m'} (z^{2 \prime})^{s-m'} \, ,
\nn
\end{align} 
and the operator $\Op$ acts entry by entry. 
The entries are equal to the Majorana polynomial of the operator $\ket{s,m} \bra{s,m'}$ written in the respective variables, $(\ket{{-\bm{n}_B'}}  \bra{-\bm{n}_B})_{m' m} = \braket{-\bm{n}_B | s, m} \braket{ s , m' |-\bm{n}_B'}$. The operator $\Op$ has to produce a Kronecker-delta $\delta_{mm'}$, which is equivalent to saying that it has to act as a trace operator on $\ket{s,m}\bra{s,m'}$. Hence, $\Op$ is similar to the trace operator \eqref{traceop},   
\begin{equation}
\Op = (N!)^{-2} ( \partial^{1'} \partial_1 + \partial^{2'} \partial_2 )^{N} \, .
\end{equation}
We can calculate the action of $\Op$ in two steps: we evaluate first the derivatives of the prime variables, yielding that
\begin{align}
&\Op (\braket{ s , m' |-\bm{n}_B'}  \braket{-\bm{n}_B | s, m} ) =  (-1)^{2s-m-m'} (N!)^{-1}
\nn
\\
&\times \sqrt{\binom{2s}{s-m} \binom{2s}{s-m'}} \pd_{1}^{s+m'} \pd_{2}^{s-m'}(  z_1^{s+m} z_2^{s-m} ) \, ,
\end{align}
 and then we let the remaining derivatives act. The last result showed us that the action of $\Op$ is equivalent to interchange the prime variables $(z^{1 \prime} , z^{2 \prime})$ by $(\partial_1 , \partial_2)$, and then we apply the remaining derivatives in the second factor of the \rhs~of Eq. \eqref{two.zetas}. $p_C(z)$ is obtained, after $\Op$ acts on \eqref{two.zetas}, by making the substitution $\bra{-\bm{n}_B'}  \rightarrow \bra{-\bm{n}_B}$, 
\begin{equation}
p_C(z)  = (N!)^{-1 } p_D(z_a , \partial_a) p_E(z_a , z^{a}) \, ,
\end{equation}
where the derivatives only act on $p_E(z)$, which can be ensured by writing the variables in each monomial of $p_D(z_a , \partial_a)$ such that the partial derivatives go to the right of the monomial, to affect only the polynomial on the right. In a similar way, we can do the same procedure evaluating first the derivatives over the variables $z_a$ instead of the prime variables $z^{a \prime}$, obtaining a similar equation as the previous one,
\begin{equation}
p_C(z)  = (N!)^{-1 } p_E( \partial^a , z^a) p_D(z_a , z^{a}) \, . \quad \square 
\end{equation}
\emph{Proof of Lemma \ref{prod.op}}
\begin{align}
(N!)^3 \Tr  \left( C D  \right)  = &\left( \pd^a \pd_a\right)^{N} \left[ 
p_{C} (z_b , \pd_b) p_{D} (z) \right]
\nn
\\
= & \left(  \pd_{c_{1} \dots c_{k}}  p_C (z_b , \pd_b) 
\right) \left( \pd^{c_{1} \dots c_{N}} \pd_{c_{k+1} \dots 
c_{2s}} p_D (z) \right)
\nn
\\
= & \left(  \pd_{c_{1} \dots c_{N}}  p_C (z_b , \pd_b) \right) \left( \pd^{c_{1} \dots c_{2s}}  p_D (z) \right)
\nn
\\
= & \left(  ( \pd^{c} \pd_{c } )^{2s}    p_C (z_b , \pd_b) 
\right) \left( p_D (z) \right)
\nn
\\
= & (N!) p_C (\pd^b , \pd_b)  \left( p_D (z) \right)
\, ,
\end{align}
where the repeated indices $c_j$ run from 1 to 2 and $\partial^{c_1 \dots c_k}$ is short notation for $\partial^{c_1} \dots \partial^{c_k}  $, and where in the second line there are no derivatives $\pd_k$ acting in $p_D(z)$, otherwise the number of partial derivatives exceeds the degree of $p_D (z)$ in the $z_a$ variables. The last equation is equivalent to the application of the operator $\Op$, and it yields the final result. $\, \square$ 
\begin{widetext}
\section{Proof of Theorem \ref{Big.Theo}}
\label{Proof.BigT}
$1.$ We use the equation \eqref{decomp.TensOp} to calculate explicitly its polynomial using \eqref{ms.pol}
\begin{align}
p_{\si \mu}(z)=& 
\sum_{m,m'} (-1)^{3s-2m'-m} C^{\si \mu}_{sm,s-m'} \sqrt{\binom{2s}{s-m}\binom{2s}{s-m'}}z_1^{s+m}z_2^{s-m}
(z^1)^{s+m'} (z^2)^{s-m'}
\, .
\label{scale.factors}
\end{align}
The action of $L$ in the last equation yields
\begin{align}
&(2s)^2 L \left( p_{\si \mu}^{(s)} \right) =
\sum_{m,m'} (-1)^{3s-2m'-m} C^{\si \mu}_{sm,s-m'} \sqrt{\binom{2s}{s-m}\binom{2s}{s-m'}} \times
\nn
\\
&\left[
(s+m)(s+m')z_1^{s+m-1}z_2^{s-m}(z^1)^{s+m'-1}(z^2)^{s-m'}
 + 
(s-m)(s-m')z_1^{s+m}z_2^{s-m-1}(z^1)^{s+m'}(z^2)^{s-m'-1}
\right]
\, ,
\nn
\\
&=
\sum_{m,m'}2s (-1)^{s-m'} C^{\si \mu}_{sm,s-m'} 
\left(
 \sqrt{(s+m)(s+m')}  \braket{-\bm{n}_B |s-1/2,m-1/2} \braket{s-1/2,m'-1/2|-\bm{n}_B}
\right.
\nn
\\
&\left.
+ \sqrt{(s-m)(s-m')} \braket{-\bm{n}_B|s-1/2,m+1/2} \braket{s-1/2,m'+1/2|-\bm{n}_B}
\right) \, ,
\nn
\\
&= 2s \sum_{m,m'}(-1)^{s-m'} 
\braket{-\bm{n}_B|s-1/2,m-1/2} \braket{s-1/2,m'-1/2|-\bm{n}_B} \times
\nn
\\
&\left(  C^{\si \mu}_{sm,s-m'}  \sqrt{(s+m)(s+m')}
- C^{\si \mu}_{sm-1,s-m'+1}  \sqrt{(s-m+1)(s-m'+1)}
 \right) 
\nn
 \\
&= 2s \sqrt{(2s-\si)(2s+\si+1)} \sum_{m,m'}(-1)^{s-m'} C^{\si \mu}_{s-1/2m-1/2,s-1/2-m'+1/2}  \times
\nn
\\
& \braket{-\bm{n}_B|s-1/2,m-1/2} \braket{s-1/2,m'-1/2|-\bm{n}_B}
\nn
\\
& =2s \sqrt{(2s-\si)(2s+\si+1)} p_{\si \mu}^{(s-1/2)} \, ,
\end{align}
where
\begin{equation}
\braket{-\bm{n}_B| s,m}= (-1)^{s-m} \sqrt{\binom{2s}{s-m}} z_1^{s+m} z_2^{s-m} \, , \quad \quad
\braket{s,m'|-\bm{n}_B}= (-1)^{s-m'} \sqrt{\binom{2s}{s-m'}} (z^1)^{s+m'} (z^2)^{s-m'} \, ,
\end{equation}
and we use the following properties of the Clebsh-Gordan coefficients (\cite{Var.Mos.Khe:88}, p.254).
\begin{align}
(2s+1)(s+m')^{1/2}C^{\si \mu}_{sm,s-m'} = &
\left[ (s+m)(2s-\si)(2s+\si+1) \right]^{1/2} C^{\si \mu}_{s-1/2m-1/2,s-1/2-m'+1/2}
\nn
\\
&+ \left[ (s-m+1)\si (\si+1) \right]^{1/2} C^{\si \mu}_{s+1/2 m-1/2,s-1/2-m'+1/2} \, ,
\\
(2s+1)(s-m'+1)^{1/2}C^{\si \mu}_{sm-1,s-m'+1} = &
-\left[ (s-m+1)(2s-\si)(2s+\si+1) \right]^{1/2} C^{\si \mu}_{s-1/2m-1/2,s-1/2-m'+1/2}
\nn
\\
&+ \left[ (s+m)\si (\si+1) \right]^{1/2} C^{\si \mu}_{s+1/2 m-1/2,s-1/2-m'+1/2} \, .
\end{align}
\end{widetext}
In particular, $L(p^{(\si/2)}_{\si \mu}(z))=0$. Now, for $p^{(s)}_{00}(z) = (2s+1)^{-1/2}(z_az^a)^{2s}$, $\sqrt{{2s}} L(p^{(s)}_{00}(z)) = \sqrt{2s+1}  p^{(s-1/2)}_{00}(z)$, or equivalent, $(2s+1)^{-1/2} T^{(s)}_{00} \rightarrow_L (2s)^{-1/2} T^{(s-1/2)}_{00} $, both of them with unit trace. Because $T_{00}^{(s)}$ is the only non-traceless operator in the basis $\{ T_{\si \mu}^{(s)} \}_{\si\,\mu}$ for each $(s)$, we conclude that the partial trace operator preserves the trace. $\square$

$2.$ The set $\{ p^{(s)}_{\si \mu}(z)\}_{\si \, \mu}$ of $P^{(N,N)}(z)$ is an orthonormal basis due to its bijection with the tensor operators $\{T^{(s)}_{\si \mu} \}_{\si \mu}$, and hence $ (z_a z^a)^{2s-\si}  p^{(\si /2)}_{\si \mu}(z) = \sum_{\tau \nu} c_{\tau \nu} p^{(s)}_{\tau \nu}(z)$ where the coefficients $c_{\tau \nu}$ can be calculated using Lemma \ref{prod.op}
\begin{align}
c_{\tau \nu} = & (-1)^{\nu} (N!)^{-2} (\pd_a \pd^a)^{N-\si}p^{(\si /2)}_{\si \mu} (\pd^a , \pd_a) \left(p_{\tau -\nu}^{(s)}(z_a , z^a) \right)
\nn
\\
\propto & (-1)^{\nu} p^{(\si /2)}_{\si \mu} (\pd^a , \pd_a) \left( p_{\tau -\nu}^{(\si/2)}(z_a , z^a) \right)
\nn
\\
\propto & \Tr (T^{(\si/2)}_{\si \mu} T^{\da (\si/2)}_{\tau \nu}) = \delta_{\si \tau} \delta_{\mu \nu} 
\, ,
\end{align}
implying that $p^{(s)}_{\si \mu} (z) = K (z_a z^a)^{2s-\si}  p^{(\si /2)}_{\si \mu}(z) $, with $K$ a proportional factor. Using Eq. \eqref{factor.L} of Theorem \ref{Big.Theo} $(2s-\si)$ times, we obtain that
\begin{widetext}
\begin{align}
L^{(2s-\si)}(p^{(s)}_{\si \mu}(z)) = & K\frac{(2s+\si+1)(2s-\si)}{(2s)^2} L^{(2s-\si-1)}\left((z^a z_a)^{2s-\si-1}p^{(\si/2)}_{\si \mu} (z) \right)
= K l(s,\si)^2 p^{(\si/2)}_{\si \mu} (z) \, ,
\end{align}
\end{widetext}
where we conclude that $K=l(s , \, \si)^{-1}$.
\section{$T_{\si \mu}^{(s)}$ in the $S$-rep}
\label{App.TScon}
In this appendix, we prove the equations \eqref{exp.TSconnection}-\eqref{factor.TSconnection}. First, we calculate eq. \eqref{exp.TSconnection} with $s=\tau/2$ and $\si= \tau$. The next equation (from \cite{Bie.Lou:81}, p.90) helps us to write the expansion of $T_{\tau \mu}^{(\tau/2)}$ in terms of the $S$ operators with $\nu_0 = 0$,
\begin{align}
T_{\tau \mu}^{(\tau/2)} = \left[ \frac{(\tau+\mu)!}{(2\tau)!(\tau-\mu)!} \right]^{1/2} 
\left[ S_-, T^{(\tau/2)}_{\tau \tau} \right]_{(\tau-\mu)} \, ,
\label{Bied.comm}
\end{align}
where $S_-$ is the ladder operator in the $(\tau/2)$-irrep, and
\begin{align}
[A,B]_{q} \equiv  \underbrace{[A,[A,\dots ,[A,B]] \dots]}_{q} \, ,
\end{align}
is the nested commutator. The operators $S_-$ and $T_{\tau \tau}^{(\tau/2)}$ in terms of the $S$-rep are equal to
\begin{align}
T_{\tau \tau}^{(\tau/2)} =& (-2)^{-\tau} S_{(0,0,0,\tau)}
\nn
\\
=& (-2)^{-\tau} \CP_{\tau/2} ( \underbrace{\si_+ \otimes \dots \otimes \si_+}_{\tau} )
\CP_{\tau/2} \, ,
\label{T.sisi}
\\
S_- = & \frac{\tau}{2} S_{(\tau-1,1,0,0)} \, ,
\nn
\\
=& \frac{\tau}{2}  \CP_{\tau/2} ( \si_- \otimes \underbrace{\si_0 \otimes \dots \otimes \si_0 }_{\tau-1} ) \CP_{\tau/2} \, .
\end{align}
The commutator in eq. \eqref{Bied.comm} can be calculated with the following
\begin{lemma}
 Let $p(z) \in P^{(N,N)}(z)$, hence
\begin{align}
1.) \, \, \, \quad z^{c_1} \dots z^{c_k} \partial^{c_1 \dots c_k} p(z) =& \frac{N!}{(N-k)!} p(z) \, ,
\\
2.) \, z^{c_1} \dots z^{c_k} z^a \partial^{c_1 \dots c_k b} p(z) =& \frac{(N-1)!}{(N-k)!}  z^a \partial^b p(z) \, ,
\end{align}
where $\partial^{c_1 \dots c_k}$ is short notation for $\partial^{c_1} \dots \partial^{c_{k}} $ and repeated indices run from 1 to 2.
\end{lemma}
\emph{Proof.}
1.) By induction: $k=2$ is easy to prove. Let us assume the result is valid for $k$ and prove it for $k+1$,
\begin{align}
&z^{c_1} \dots z^{c_{k+1}}  \partial^{c_1 \dots c_{k+1} } p(z) 
\nn
\\
=& z^{c_1} \dots z^{c_{k}} \partial^{c_1} \left(  z^{c_{k+1}} \partial^{c_2 \dots c_{k+1} } p(z)  \right)
\nn
\\
& - z^{c_{k+1}} z^{c_2} \dots z^{c_{k}} \partial^{c_2 \dots c_{k+1} } p(z) 
\nn
\\
=& z^{c_1} \dots z^{c_{k}} \partial^{c_1 \dots c_k} \left(  z^{c_{k+1}} \partial^{ c_{k+1} } ( p(z) ) - k p(z)  \right)
\nn
\\
=&  \frac{N!}{(N-(k+1))!} p(z)
\, .
\end{align}
The proof of 2.) is analogue. $\square$

The commutator $G\equiv [S_{-} , S_{\vec{\nu}}]$ is calculated with polynomials using the previous result, Lemma \ref{Action.optoop} and eq. \eqref{fac.pol},
\begin{align}
p_G(z) =& (\tau!)^{-1} (p_{S_-} (z_a, \partial_a ) - p_{S_-} (\partial^a , z^a) ) p_{S_{\vec{\nu}}}(z)
\nn
\\
=& \frac{1}{2} (p_{-} (z_a ,\partial_a) - p_{-} (\partial^a, z^a))
\prod_j (p_j(z))^{\nu_j} 
\\
=&(p_0)^{\nu_0} (p_-)^{\nu_-} (p_z)^{\nu_z-1} (p_+)^{\nu_+-1} (\nu_z p_- p_+ - 2\nu_+ p_z^2 ) \, ,
\nn
\end{align}
where we use the commutators of the Pauli matrices and ladder operators
\begin{align}
[\si_- , \si_z]= 2\si_- , \, \qquad [\si_- , \si_+]= -4\si_z \, .
\end{align}
We obtain that
\begin{equation}
[S_{-} , S_{\vec{\nu}}] = \nu_z S_{(\nu_0 , \, \nu_-+1 , \, \nu_z-1 , \nu_+)} - 2 \nu_+ S_{(\nu_0 , \, \nu_- , \, \nu_z+1 , \nu_+-1)} \, .
\label{comm.S}
\end{equation}
The constants $\nu_z$ and $\nu_+$ can be thought as the number of possible operators $\si_z$ and $\si_+$ where one can apply the commutator of $\si_-$. The next step to do is the calculation of the equation \eqref{Bied.comm} applying iteratively the latter result. This implies that $T_{\tau \mu}^{(\tau/2)}$ is a linear combination of $S$-operators satisfying the following: $\nu_0=0$, $\nu_+ - \nu_- = \mu$ and $\nu_- + \nu_z + \nu_+= \tau$. Hence
\begin{equation}
T_{\tau \mu}^{(\tau/2)} = 
 \sum_{k = \mu}^{\tau} 
 A(\tau/2,  \tau ,  \mu ,  k)  S_{(0, k -\mu, \tau + \mu - 2k , k)}  \, ,
\label{TS.nu0}
\end{equation}
with the condition that $\tau+\mu-2k \geq0$. $ A(\tau/2,  \tau ,  \mu ,  k)$ accumulates the constant factors of eqs. \eqref{Bied.comm}, \eqref{T.sisi}, the factor $(-2)^{\tau-k}$ from eq. \eqref{comm.S}, where the exponent is the difference between the initial and final values of $\nu_+$, and a combinatorial number given by: the number of ways to choose $(\tau-k)$ $\si_+$ operators from a set of $\tau$, $\binom{\tau}{\tau-k}$ (to apply $[\si_-, \bullet]$ and obtain $\si_z$); the number of ways to choose $(k-\mu)$ $\si_z$ operators from a set of $(\tau-k)$, $\binom{\tau-k}{k-\mu}$ (to apply $[\si_-, \bullet]$ and obtain $\si_-$); and the number of possible orders to apply the $(\tau-\mu)$ $\si_-$ operators to obtain the respective operator $S_{(0, k -\mu, \tau + \mu - 2k , k)}$, $(\tau-\mu)!/2^{k-\mu}$. The expression of $A(\tau/2, \tau, \mu,k) $ is equal to 
\begin{widetext}
\begin{equation}
A\left(\frac{\tau}{2}, \tau, \mu,k \right) = 
\sqrt{
\frac{(\tau+\mu)!}{(2\tau)!(\tau-\mu)!} 
}
(-2)^{-k} \binom{\tau}{\tau-k}
\binom{\tau-k}{k-\mu} \frac{(\tau-\mu)!}{2^{k-\mu}}
=
\sqrt{
\frac{(\tau+\mu)!(\tau-\mu)!}{(2\tau)!} 
}
\frac{(-1)^{k} 2^{\mu-2k}(\tau!)}{ k! (k-\mu)! (\tau+\mu-2k)!} \, .
\end{equation}
\end{widetext}
The equations \eqref{exp.TSconnection}-\eqref{factor.TSconnection} for $T^{(s)}_{\si \mu}$ and $A(s, \, \si, \, \mu , \, k)$ for a general $s$ is obtained with the polynomial expression of eq. \eqref{TS.nu0} after we multiply by $(z^az_a)^{2s-\si}$ and use Theorem \ref{Big.Theo}.
\end{appendix}
%
\bibliography{mybibs_bt}

\begin{thebibliography}{54}%
\makeatletter
\providecommand \@ifxundefined [1]{%
 \@ifx{#1\undefined}
}%
\providecommand \@ifnum [1]{%
 \ifnum #1\expandafter \@firstoftwo
 \else \expandafter \@secondoftwo
 \fi
}%
\providecommand \@ifx [1]{%
 \ifx #1\expandafter \@firstoftwo
 \else \expandafter \@secondoftwo
 \fi
}%
\providecommand \natexlab [1]{#1}%
\providecommand \enquote  [1]{``#1''}%
\providecommand \bibnamefont  [1]{#1}%
\providecommand \bibfnamefont [1]{#1}%
\providecommand \citenamefont [1]{#1}%
\providecommand \href@noop [0]{\@secondoftwo}%
\providecommand \href [0]{\begingroup \@sanitize@url \@href}%
\providecommand \@href[1]{\@@startlink{#1}\@@href}%
\providecommand \@@href[1]{\endgroup#1\@@endlink}%
\providecommand \@sanitize@url [0]{\catcode `\\12\catcode `\$12\catcode
  `\&12\catcode `\#12\catcode `\^12\catcode `\_12\catcode `\%12\relax}%
\providecommand \@@startlink[1]{}%
\providecommand \@@endlink[0]{}%
\providecommand \url  [0]{\begingroup\@sanitize@url \@url }%
\providecommand \@url [1]{\endgroup\@href {#1}{\urlprefix }}%
\providecommand \urlprefix  [0]{URL }%
\providecommand \Eprint [0]{\href }%
\providecommand \doibase [0]{http://dx.doi.org/}%
\providecommand \selectlanguage [0]{\@gobble}%
\providecommand \bibinfo  [0]{\@secondoftwo}%
\providecommand \bibfield  [0]{\@secondoftwo}%
\providecommand \translation [1]{[#1]}%
\providecommand \BibitemOpen [0]{}%
\providecommand \bibitemStop [0]{}%
\providecommand \bibitemNoStop [0]{.\EOS\space}%
\providecommand \EOS [0]{\spacefactor3000\relax}%
\providecommand \BibitemShut  [1]{\csname bibitem#1\endcsname}%
\let\auto@bib@innerbib\@empty
\bibitem [{\citenamefont {Giraud}\ \emph {et~al.}(2015)\citenamefont {Giraud},
  \citenamefont {Braun}, \citenamefont {Baguette}, \citenamefont {Bastin},\
  and\ \citenamefont {Martin}}]{Gir.Bra.Bag.Bas.Mar:15}%
  \BibitemOpen
  \bibfield  {author} {\bibinfo {author} {\bibfnamefont {O.}~\bibnamefont
  {Giraud}}, \bibinfo {author} {\bibfnamefont {D.}~\bibnamefont {Braun}},
  \bibinfo {author} {\bibfnamefont {D.}~\bibnamefont {Baguette}}, \bibinfo
  {author} {\bibfnamefont {T.}~\bibnamefont {Bastin}}, \ and\ \bibinfo {author}
  {\bibfnamefont {J.}~\bibnamefont {Martin}},\ }\href@noop {} {\bibfield
  {journal} {\bibinfo  {journal} {Phys. Rev. Lett.}\ }\textbf {\bibinfo
  {volume} {114}},\ \bibinfo {pages} {080401} (\bibinfo {year}
  {2015})}\BibitemShut {NoStop}%
\bibitem [{\citenamefont {Majorana}(1932)}]{Majorana1932}%
  \BibitemOpen
  \bibfield  {author} {\bibinfo {author} {\bibfnamefont {E.}~\bibnamefont
  {Majorana}},\ }\href@noop {} {\bibfield  {journal} {\bibinfo  {journal}
  {Nuovo Cimento}\ }\textbf {\bibinfo {volume} {9}},\ \bibinfo {pages} {43}
  (\bibinfo {year} {1932})}\BibitemShut {NoStop}%
\bibitem [{\citenamefont {Perelomov}(1986)}]{Perelmov86}%
  \BibitemOpen
  \bibfield  {author} {\bibinfo {author} {\bibfnamefont {A.}~\bibnamefont
  {Perelomov}},\ }\href@noop {} {\emph {\bibinfo {title} {Generalzed Coherent
  States and Their Applications}}}\ (\bibinfo  {publisher} {Springer-Verlag,
  Berlin},\ \bibinfo {year} {1986})\BibitemShut {NoStop}%
\bibitem [{\citenamefont {Radcliffe}(1971)}]{Rad:71}%
  \BibitemOpen
  \bibfield  {author} {\bibinfo {author} {\bibfnamefont {J.}~\bibnamefont
  {Radcliffe}},\ }\href@noop {} {\bibfield  {journal} {\bibinfo  {journal} {J.
  Phys. A: Gen. Phys.}\ }\textbf {\bibinfo {volume} {4}},\ \bibinfo {pages}
  {313} (\bibinfo {year} {1971})}\BibitemShut {NoStop}%
\bibitem [{\citenamefont {Giraud}\ \emph {et~al.}(2010)\citenamefont {Giraud},
  \citenamefont {Braun},\ and\ \citenamefont {Braun}}]{Giraud10}%
  \BibitemOpen
  \bibfield  {author} {\bibinfo {author} {\bibfnamefont {O.}~\bibnamefont
  {Giraud}}, \bibinfo {author} {\bibfnamefont {P.}~\bibnamefont {Braun}}, \
  and\ \bibinfo {author} {\bibfnamefont {D.}~\bibnamefont {Braun}},\
  }\href@noop {} {\bibfield  {journal} {\bibinfo  {journal} {New Journal of
  Physics}\ }\textbf {\bibinfo {volume} {12}},\ \bibinfo {pages} {063005}
  (\bibinfo {year} {2010})}\BibitemShut {NoStop}%
\bibitem [{\citenamefont {{Bohnet-Waldraff}}\ \emph {et~al.}(2016)\citenamefont
  {{Bohnet-Waldraff}}, \citenamefont {Braun},\ and\ \citenamefont
  {Giraud}}]{Boh.Bra.Gir:16}%
  \BibitemOpen
  \bibfield  {author} {\bibinfo {author} {\bibfnamefont {F.}~\bibnamefont
  {{Bohnet-Waldraff}}}, \bibinfo {author} {\bibfnamefont {D.}~\bibnamefont
  {Braun}}, \ and\ \bibinfo {author} {\bibfnamefont {O.}~\bibnamefont
  {Giraud}},\ }\href {\doibase 10.1103/PhysRevA.93.012104} {\bibfield
  {journal} {\bibinfo  {journal} {Phys. Rev. A}\ }\textbf {\bibinfo {volume}
  {93}},\ \bibinfo {pages} {012104} (\bibinfo {year} {2016})}\BibitemShut
  {NoStop}%
\bibitem [{\citenamefont {Zimba}(2006)}]{Zimba06}%
  \BibitemOpen
  \bibfield  {author} {\bibinfo {author} {\bibfnamefont {J.}~\bibnamefont
  {Zimba}},\ }\href@noop {} {\bibfield  {journal} {\bibinfo  {journal} {EJTP}\
  }\textbf {\bibinfo {volume} {3}},\ \bibinfo {pages} {143} (\bibinfo {year}
  {2006})}\BibitemShut {NoStop}%
\bibitem [{\citenamefont {Baguette}\ and\ \citenamefont
  {Martin}(2017)}]{Bag.Mar:17}%
  \BibitemOpen
  \bibfield  {author} {\bibinfo {author} {\bibfnamefont {D.}~\bibnamefont
  {Baguette}}\ and\ \bibinfo {author} {\bibfnamefont {J.}~\bibnamefont
  {Martin}},\ }\href@noop {} {\bibfield  {journal} {\bibinfo  {journal}
  {Phys.{} Rev.{} A}\ }\textbf {\bibinfo {volume} {96}},\ \bibinfo {pages}
  {032304} (\bibinfo {year} {2017})}\BibitemShut {NoStop}%
\bibitem [{\citenamefont {Baguette}\ \emph {et~al.}(2015)\citenamefont
  {Baguette}, \citenamefont {Damanet}, \citenamefont {Giraud},\ and\
  \citenamefont {Martin}}]{Bag.Dam.Gir.Mar:17}%
  \BibitemOpen
  \bibfield  {author} {\bibinfo {author} {\bibfnamefont {D.}~\bibnamefont
  {Baguette}}, \bibinfo {author} {\bibfnamefont {F.}~\bibnamefont {Damanet}},
  \bibinfo {author} {\bibfnamefont {O.}~\bibnamefont {Giraud}}, \ and\ \bibinfo
  {author} {\bibfnamefont {J.}~\bibnamefont {Martin}},\ }\href@noop {}
  {\bibfield  {journal} {\bibinfo  {journal} {Phys.{} Rev.{} A}\ }\textbf
  {\bibinfo {volume} {92}},\ \bibinfo {pages} {052333} (\bibinfo {year}
  {2015})}\BibitemShut {NoStop}%
\bibitem [{\citenamefont {de~la Hoz}\ \emph {et~al.}(2013)\citenamefont {de~la
  Hoz}, \citenamefont {Klimov}, \citenamefont {Kim}, \citenamefont
  {M\"{u}ller}, \citenamefont {Marquardt}, \citenamefont {Leuchs},\ and\
  \citenamefont {S\'{a}nchez-Soto}}]{Hoz.Kli.Bjo:13}%
  \BibitemOpen
  \bibfield  {author} {\bibinfo {author} {\bibfnamefont {P.}~\bibnamefont
  {de~la Hoz}}, \bibinfo {author} {\bibfnamefont {A.~B.}\ \bibnamefont
  {Klimov}}, \bibinfo {author} {\bibfnamefont {Y.-H.}\ \bibnamefont {Kim}},
  \bibinfo {author} {\bibfnamefont {C.}~\bibnamefont {M\"{u}ller}}, \bibinfo
  {author} {\bibfnamefont {C.}~\bibnamefont {Marquardt}}, \bibinfo {author}
  {\bibfnamefont {G.}~\bibnamefont {Leuchs}}, \ and\ \bibinfo {author}
  {\bibfnamefont {L.~L.}\ \bibnamefont {S\'{a}nchez-Soto}},\ }\href@noop {}
  {\bibfield  {journal} {\bibinfo  {journal} {Phys. Rev. A}\ }\textbf {\bibinfo
  {volume} {88}},\ \bibinfo {pages} {063803} (\bibinfo {year}
  {2013})}\BibitemShut {NoStop}%
\bibitem [{\citenamefont {Baecklund}\ and\ \citenamefont
  {Bengtsson}(2014)}]{Bae.Ing:14}%
  \BibitemOpen
  \bibfield  {author} {\bibinfo {author} {\bibfnamefont {A.}~\bibnamefont
  {Baecklund}}\ and\ \bibinfo {author} {\bibfnamefont {I.}~\bibnamefont
  {Bengtsson}},\ }\href@noop {} {\bibfield  {journal} {\bibinfo  {journal}
  {Phys. Scr.}\ }\textbf {\bibinfo {volume} {T163}},\ \bibinfo {pages} {014012}
  (\bibinfo {year} {2014})}\BibitemShut {NoStop}%
\bibitem [{\citenamefont {Kolenderski}\ and\ \citenamefont
  {Demkowicz-Dobrzanski}(2008)}]{Kolenderski08}%
  \BibitemOpen
  \bibfield  {author} {\bibinfo {author} {\bibfnamefont {P.}~\bibnamefont
  {Kolenderski}}\ and\ \bibinfo {author} {\bibfnamefont {R.}~\bibnamefont
  {Demkowicz-Dobrzanski}},\ }\href {\doibase 10.1103/PhysRevA.78.052333}
  {\bibfield  {journal} {\bibinfo  {journal} {Phys. Rev. A}\ }\textbf {\bibinfo
  {volume} {78}},\ \bibinfo {pages} {052333} (\bibinfo {year}
  {2008})}\BibitemShut {NoStop}%
\bibitem [{\citenamefont {Bouchard}\ \emph {et~al.}(2017)\citenamefont
  {Bouchard} \emph {et~al.}}]{Bou.etal:17}%
  \BibitemOpen
  \bibfield  {author} {\bibinfo {author} {\bibfnamefont {F.}~\bibnamefont
  {Bouchard}} \emph {et~al.},\ }\href@noop {} {\bibfield  {journal} {\bibinfo
  {journal} {Optica}\ }\textbf {\bibinfo {volume} {4}},\ \bibinfo {pages}
  {1429} (\bibinfo {year} {2017})}\BibitemShut {NoStop}%
\bibitem [{\citenamefont {Chryssomalakos}\ and\ \citenamefont
  {Hern\'{a}ndez-Coronado}(2017)}]{Chr.Her:17}%
  \BibitemOpen
  \bibfield  {author} {\bibinfo {author} {\bibfnamefont {C.}~\bibnamefont
  {Chryssomalakos}}\ and\ \bibinfo {author} {\bibfnamefont {H.}~\bibnamefont
  {Hern\'{a}ndez-Coronado}},\ }\href@noop {} {\bibfield  {journal} {\bibinfo
  {journal} {Phys.{} Rev.{} A.}\ }\textbf {\bibinfo {volume} {95}},\ \bibinfo
  {pages} {052125} (\bibinfo {year} {2017})}\BibitemShut {NoStop}%
\bibitem [{\citenamefont {Goldberg}\ and\ \citenamefont
  {James}(2018)}]{Gol.Jam:18}%
  \BibitemOpen
  \bibfield  {author} {\bibinfo {author} {\bibfnamefont {A.~Z.}\ \bibnamefont
  {Goldberg}}\ and\ \bibinfo {author} {\bibfnamefont {F.~V.}\ \bibnamefont
  {James}},\ }\href@noop {} {\bibfield  {journal} {\bibinfo  {journal} {Phys.
  Rev. A}\ }\textbf {\bibinfo {volume} {98}},\ \bibinfo {pages} {032113}
  (\bibinfo {year} {2018})}\BibitemShut {NoStop}%
\bibitem [{\citenamefont {I.Bengtsson}\ and\ \citenamefont
  {K.\.{Z}yczkowski}(2017)}]{Bengtsson17}%
  \BibitemOpen
  \bibfield  {author} {\bibinfo {author} {\bibnamefont {I.Bengtsson}}\ and\
  \bibinfo {author} {\bibnamefont {K.\.{Z}yczkowski}},\ }\href@noop {} {\emph
  {\bibinfo {title} {Geometry of quantum states: an introduction to quantum
  entanglement}}}\ (\bibinfo  {publisher} {Cambride University Press},\
  \bibinfo {year} {2017})\ \bibinfo {note} {2nd. Edition}\BibitemShut {NoStop}%
\bibitem [{\citenamefont {Barnett}\ \emph {et~al.}(2006)\citenamefont
  {Barnett}, \citenamefont {Turner},\ and\ \citenamefont
  {Demler}}]{Bar.Tur.Dem:06}%
  \BibitemOpen
  \bibfield  {author} {\bibinfo {author} {\bibfnamefont {R.}~\bibnamefont
  {Barnett}}, \bibinfo {author} {\bibfnamefont {A.}~\bibnamefont {Turner}}, \
  and\ \bibinfo {author} {\bibfnamefont {E.}~\bibnamefont {Demler}},\
  }\href@noop {} {\bibfield  {journal} {\bibinfo  {journal} {Phys. Rev. Lett.}\
  }\textbf {\bibinfo {volume} {97}},\ \bibinfo {pages} {180412} (\bibinfo
  {year} {2006})}\BibitemShut {NoStop}%
\bibitem [{\citenamefont {Barnett}\ \emph {et~al.}(2007)\citenamefont
  {Barnett}, \citenamefont {Turner},\ and\ \citenamefont
  {Demler}}]{Bar.Tur.Dem:07}%
  \BibitemOpen
  \bibfield  {author} {\bibinfo {author} {\bibfnamefont {R.}~\bibnamefont
  {Barnett}}, \bibinfo {author} {\bibfnamefont {A.}~\bibnamefont {Turner}}, \
  and\ \bibinfo {author} {\bibfnamefont {E.}~\bibnamefont {Demler}},\
  }\href@noop {} {\bibfield  {journal} {\bibinfo  {journal} {Phys. Rev. Lett.}\
  }\textbf {\bibinfo {volume} {76}},\ \bibinfo {pages} {013605} (\bibinfo
  {year} {2007})}\BibitemShut {NoStop}%
\bibitem [{\citenamefont {M{\"a}kel{\"a}}\ and\ \citenamefont
  {Suominen}(2007)}]{Mak.Suo:07}%
  \BibitemOpen
  \bibfield  {author} {\bibinfo {author} {\bibfnamefont {H.}~\bibnamefont
  {M{\"a}kel{\"a}}}\ and\ \bibinfo {author} {\bibfnamefont {K.~A.}\
  \bibnamefont {Suominen}},\ }\href@noop {} {\bibfield  {journal} {\bibinfo
  {journal} {Phys.Rev.Lett.}\ }\textbf {\bibinfo {volume} {99}},\ \bibinfo
  {pages} {190408} (\bibinfo {year} {2007})}\BibitemShut {NoStop}%
\bibitem [{\citenamefont {Ribeiro}\ \emph {et~al.}(2007)\citenamefont
  {Ribeiro}, \citenamefont {Vidal},\ and\ \citenamefont
  {Mosseri}}]{Rib.Vid.Mos:07}%
  \BibitemOpen
  \bibfield  {author} {\bibinfo {author} {\bibfnamefont {P.}~\bibnamefont
  {Ribeiro}}, \bibinfo {author} {\bibfnamefont {J.}~\bibnamefont {Vidal}}, \
  and\ \bibinfo {author} {\bibfnamefont {R.}~\bibnamefont {Mosseri}},\
  }\href@noop {} {\bibfield  {journal} {\bibinfo  {journal} {Phys.{} Rev.{}
  Lett.}\ }\textbf {\bibinfo {volume} {99}},\ \bibinfo {pages} {050402}
  (\bibinfo {year} {2007})}\BibitemShut {NoStop}%
\bibitem [{\citenamefont {Ribeiro}\ \emph {et~al.}(2008)\citenamefont
  {Ribeiro}, \citenamefont {Vidal},\ and\ \citenamefont
  {Mosseri}}]{Rib.Vid.Mos:08}%
  \BibitemOpen
  \bibfield  {author} {\bibinfo {author} {\bibfnamefont {P.}~\bibnamefont
  {Ribeiro}}, \bibinfo {author} {\bibfnamefont {J.}~\bibnamefont {Vidal}}, \
  and\ \bibinfo {author} {\bibfnamefont {R.}~\bibnamefont {Mosseri}},\
  }\href@noop {} {\bibfield  {journal} {\bibinfo  {journal} {Phys.{} Rev.{} E}\
  }\textbf {\bibinfo {volume} {78}},\ \bibinfo {pages} {021106} (\bibinfo
  {year} {2008})}\BibitemShut {NoStop}%
\bibitem [{\citenamefont {Atiyah}(2001)}]{Ati:01}%
  \BibitemOpen
  \bibfield  {author} {\bibinfo {author} {\bibfnamefont {M.}~\bibnamefont
  {Atiyah}},\ }\href@noop {} {\bibfield  {journal} {\bibinfo  {journal} {Phil.
  Trans. R. Soc. Lond. A}\ }\textbf {\bibinfo {volume} {359}},\ \bibinfo
  {pages} {1} (\bibinfo {year} {2001})}\BibitemShut {NoStop}%
\bibitem [{\citenamefont {Leboeuf}\ and\ \citenamefont
  {Voros}(1990)}]{Leb.Vor:90}%
  \BibitemOpen
  \bibfield  {author} {\bibinfo {author} {\bibfnamefont {P.}~\bibnamefont
  {Leboeuf}}\ and\ \bibinfo {author} {\bibfnamefont {A.}~\bibnamefont
  {Voros}},\ }\href@noop {} {\bibfield  {journal} {\bibinfo  {journal} {J.
  Phys. A: Math. Gen.}\ }\textbf {\bibinfo {volume} {23}},\ \bibinfo {pages}
  {1765} (\bibinfo {year} {1990})}\BibitemShut {NoStop}%
\bibitem [{\citenamefont {Haldane}\ and\ \citenamefont
  {Rezayi}(1985)}]{Hal.Rey:85}%
  \BibitemOpen
  \bibfield  {author} {\bibinfo {author} {\bibfnamefont {F.~D.}\ \bibnamefont
  {Haldane}}\ and\ \bibinfo {author} {\bibfnamefont {E.~H.}\ \bibnamefont
  {Rezayi}},\ }\href@noop {} {\bibfield  {journal} {\bibinfo  {journal} {Phys.
  Rev. B}\ }\textbf {\bibinfo {volume} {31}},\ \bibinfo {pages} {2529}
  (\bibinfo {year} {1985})}\BibitemShut {NoStop}%
\bibitem [{\citenamefont {Arovas}\ \emph {et~al.}(1988)\citenamefont {Arovas},
  \citenamefont {Bhatt}, \citenamefont {Haldane}, \citenamefont {Littlewood},\
  and\ \citenamefont {Rammal}}]{Aro.Bha.Hal.Lit.Ram:88}%
  \BibitemOpen
  \bibfield  {author} {\bibinfo {author} {\bibfnamefont {D.~P.}\ \bibnamefont
  {Arovas}}, \bibinfo {author} {\bibfnamefont {R.~N.}\ \bibnamefont {Bhatt}},
  \bibinfo {author} {\bibfnamefont {F.~D.~M.}\ \bibnamefont {Haldane}},
  \bibinfo {author} {\bibfnamefont {P.}~\bibnamefont {Littlewood}}, \ and\
  \bibinfo {author} {\bibfnamefont {R.}~\bibnamefont {Rammal}},\ }\href@noop {}
  {\bibfield  {journal} {\bibinfo  {journal} {Phys. Rev. B}\ }\textbf {\bibinfo
  {volume} {60}},\ \bibinfo {pages} {619} (\bibinfo {year} {1988})}\BibitemShut
  {NoStop}%
\bibitem [{\citenamefont {Byrd}\ and\ \citenamefont
  {Khaneja}(2003)}]{Byr.Kha:03}%
  \BibitemOpen
  \bibfield  {author} {\bibinfo {author} {\bibfnamefont {M.}~\bibnamefont
  {Byrd}}\ and\ \bibinfo {author} {\bibfnamefont {N.}~\bibnamefont {Khaneja}},\
  }\href@noop {} {\bibfield  {journal} {\bibinfo  {journal} {Phys. Rev. A}\
  }\textbf {\bibinfo {volume} {68}},\ \bibinfo {pages} {062322} (\bibinfo
  {year} {2003})}\BibitemShut {NoStop}%
\bibitem [{\citenamefont {M{\"a}kel{\"a}}\ and\ \citenamefont
  {Messina}(2010{\natexlab{a}})}]{Mak.Mes:10}%
  \BibitemOpen
  \bibfield  {author} {\bibinfo {author} {\bibfnamefont {H.}~\bibnamefont
  {M{\"a}kel{\"a}}}\ and\ \bibinfo {author} {\bibfnamefont {A.}~\bibnamefont
  {Messina}},\ }\href@noop {} {\bibfield  {journal} {\bibinfo  {journal} {Phys.
  Rev. A}\ }\textbf {\bibinfo {volume} {81}},\ \bibinfo {pages} {012326}
  (\bibinfo {year} {2010}{\natexlab{a}})}\BibitemShut {NoStop}%
\bibitem [{\citenamefont {M{\"a}kel{\"a}}\ and\ \citenamefont
  {Messina}(2010{\natexlab{b}})}]{Mak.Mes:10b}%
  \BibitemOpen
  \bibfield  {author} {\bibinfo {author} {\bibfnamefont {H.}~\bibnamefont
  {M{\"a}kel{\"a}}}\ and\ \bibinfo {author} {\bibfnamefont {A.}~\bibnamefont
  {Messina}},\ }\href@noop {} {\bibfield  {journal} {\bibinfo  {journal} {Phys.
  Scr.}\ }\textbf {\bibinfo {volume} {2010}},\ \bibinfo {pages} {014054}
  (\bibinfo {year} {2010}{\natexlab{b}})}\BibitemShut {NoStop}%
\bibitem [{\citenamefont {Aerts}\ and\ \citenamefont
  {de~Bianchi}(2016)}]{Aer.Bia:16}%
  \BibitemOpen
  \bibfield  {author} {\bibinfo {author} {\bibfnamefont {D.}~\bibnamefont
  {Aerts}}\ and\ \bibinfo {author} {\bibfnamefont {M.~S.}\ \bibnamefont
  {de~Bianchi}},\ }\href@noop {} {\bibfield  {journal} {\bibinfo  {journal} {J.
  Math. Phys.}\ }\textbf {\bibinfo {volume} {57}},\ \bibinfo {pages} {122110}
  (\bibinfo {year} {2016})}\BibitemShut {NoStop}%
\bibitem [{\citenamefont {Br\"{u}ning}\ \emph {et~al.}(2012)\citenamefont
  {Br\"{u}ning}, \citenamefont {M\"{a}kel\"{a}}, \citenamefont {Messina},\ and\
  \citenamefont {Petruccione}}]{Bru.Mak.Mes.Pet:12}%
  \BibitemOpen
  \bibfield  {author} {\bibinfo {author} {\bibfnamefont {E.}~\bibnamefont
  {Br\"{u}ning}}, \bibinfo {author} {\bibfnamefont {H.}~\bibnamefont
  {M\"{a}kel\"{a}}}, \bibinfo {author} {\bibfnamefont {A.}~\bibnamefont
  {Messina}}, \ and\ \bibinfo {author} {\bibfnamefont {F.}~\bibnamefont
  {Petruccione}},\ }\href@noop {} {\bibfield  {journal} {\bibinfo  {journal}
  {J. Mod. Opt.}\ }\textbf {\bibinfo {volume} {59}},\ \bibinfo {pages} {1}
  (\bibinfo {year} {2012})}\BibitemShut {NoStop}%
\bibitem [{\citenamefont {Ashourisheikhi}\ and\ \citenamefont
  {Sirsi}(2013)}]{Ash.Sir:13}%
  \BibitemOpen
  \bibfield  {author} {\bibinfo {author} {\bibfnamefont {S.}~\bibnamefont
  {Ashourisheikhi}}\ and\ \bibinfo {author} {\bibfnamefont {S.}~\bibnamefont
  {Sirsi}},\ }\href@noop {} {\bibfield  {journal} {\bibinfo  {journal} {Int. J.
  Quantum Inf.}\ }\textbf {\bibinfo {volume} {11}},\ \bibinfo {pages} {1350072}
  (\bibinfo {year} {2013})}\BibitemShut {NoStop}%
\bibitem [{\citenamefont {Giraud}\ \emph {et~al.}(2012)\citenamefont {Giraud},
  \citenamefont {Braun},\ and\ \citenamefont
  {Braun}}]{giraud_parametrization_2012}%
  \BibitemOpen
  \bibfield  {author} {\bibinfo {author} {\bibfnamefont {O.}~\bibnamefont
  {Giraud}}, \bibinfo {author} {\bibfnamefont {P.}~\bibnamefont {Braun}}, \
  and\ \bibinfo {author} {\bibfnamefont {D.}~\bibnamefont {Braun}},\ }\href
  {\doibase 10.1103/PhysRevA.85.032101} {\bibfield  {journal} {\bibinfo
  {journal} {Phys. Rev. A}\ }\textbf {\bibinfo {volume} {85}},\ \bibinfo
  {pages} {032101} (\bibinfo {year} {2012})}\BibitemShut {NoStop}%
\bibitem [{\citenamefont {Caban}\ \emph {et~al.}(2015)\citenamefont {Caban},
  \citenamefont {Rembielinski}, \citenamefont {Smolinski},\ and\ \citenamefont
  {Walczak}}]{Cab.Rem.Smo.Wal:15}%
  \BibitemOpen
  \bibfield  {author} {\bibinfo {author} {\bibfnamefont {P.}~\bibnamefont
  {Caban}}, \bibinfo {author} {\bibfnamefont {J.}~\bibnamefont {Rembielinski}},
  \bibinfo {author} {\bibfnamefont {.~K.}\ \bibnamefont {Smolinski}}, \ and\
  \bibinfo {author} {\bibfnamefont {Z.}~\bibnamefont {Walczak}},\ }\href@noop
  {} {\bibfield  {journal} {\bibinfo  {journal} {Quantum Inf. Process.}\
  }\textbf {\bibinfo {volume} {14}},\ \bibinfo {pages} {4665} (\bibinfo {year}
  {2015})}\BibitemShut {NoStop}%
\bibitem [{Note1()}]{Note1}%
  \BibitemOpen
  \bibinfo {note} {We say the projective space of polynomials instead of the
  projectivization of a polynomial ring because we do not consider the
  multiplication operation of polynomials.}\BibitemShut {Stop}%
\bibitem [{\citenamefont {Ramachandran}\ and\ \citenamefont
  {Ravishankar}(1986)}]{Ram.Rav:86}%
  \BibitemOpen
  \bibfield  {author} {\bibinfo {author} {\bibfnamefont {G.}~\bibnamefont
  {Ramachandran}}\ and\ \bibinfo {author} {\bibfnamefont {V.}~\bibnamefont
  {Ravishankar}},\ }\href@noop {} {\bibfield  {journal} {\bibinfo  {journal}
  {J. Phys. G: Nucl. Phys.}\ }\textbf {\bibinfo {volume} {12}},\ \bibinfo
  {pages} {L143} (\bibinfo {year} {1986})}\BibitemShut {NoStop}%
\bibitem [{Note2()}]{Note2}%
  \BibitemOpen
  \bibinfo {note} {This representation has been called the Multiaxial
  representation (MAR) in other works \cite {Ash.Sir:13, Sum.Sir.Hed.Bha:17}.
  However, we think that it is a confusing name because usually axes are
  unoriented objects while, as it is explained here and in the original work
  \cite {Ram.Rav:86}, the orientation of the axes provides relevant information
  to uniquely specify the state. If one does not consider the axes orientations
  (or subconstellations) as was done in \cite {Ash.Sir:13, Sum.Sir.Hed.Bha:17},
  the mapping from the states to the MAR representation is not
  injective.}\BibitemShut {Stop}%
\bibitem [{\citenamefont {Bohnet-Waldraff}\ \emph {et~al.}(2016)\citenamefont
  {Bohnet-Waldraff}, \citenamefont {Braun},\ and\ \citenamefont
  {Giraud}}]{boh.bra.gir:16.2}%
  \BibitemOpen
  \bibfield  {author} {\bibinfo {author} {\bibfnamefont {F.}~\bibnamefont
  {Bohnet-Waldraff}}, \bibinfo {author} {\bibfnamefont {D.}~\bibnamefont
  {Braun}}, \ and\ \bibinfo {author} {\bibfnamefont {O.}~\bibnamefont
  {Giraud}},\ }\href@noop {} {\bibfield  {journal} {\bibinfo  {journal} {Phys.
  Rev. A}\ }\textbf {\bibinfo {volume} {94}},\ \bibinfo {pages} {042343}
  (\bibinfo {year} {2016})}\BibitemShut {NoStop}%
\bibitem [{\citenamefont {Bohnet-Waldraff}\ \emph {et~al.}(2017)\citenamefont
  {Bohnet-Waldraff}, \citenamefont {Giraud},\ and\ \citenamefont
  {Braun}}]{Boh.Gir.Bra:17}%
  \BibitemOpen
  \bibfield  {author} {\bibinfo {author} {\bibfnamefont {F.}~\bibnamefont
  {Bohnet-Waldraff}}, \bibinfo {author} {\bibfnamefont {O.}~\bibnamefont
  {Giraud}}, \ and\ \bibinfo {author} {\bibfnamefont {D.}~\bibnamefont
  {Braun}},\ }\href@noop {} {\bibfield  {journal} {\bibinfo  {journal} {Phys.
  Rev. A}\ }\textbf {\bibinfo {volume} {95}},\ \bibinfo {pages} {012318}
  (\bibinfo {year} {2017})}\BibitemShut {NoStop}%
\bibitem [{\citenamefont {Designolle}\ \emph {et~al.}(2017)\citenamefont
  {Designolle}, \citenamefont {Giraud},\ and\ \citenamefont
  {Martin}}]{Des.Gir.Mar:17}%
  \BibitemOpen
  \bibfield  {author} {\bibinfo {author} {\bibfnamefont {S.}~\bibnamefont
  {Designolle}}, \bibinfo {author} {\bibfnamefont {O.}~\bibnamefont {Giraud}},
  \ and\ \bibinfo {author} {\bibfnamefont {J.}~\bibnamefont {Martin}},\ }\href
  {\doibase 10.1103/PhysRevA.96.032322} {\bibfield  {journal} {\bibinfo
  {journal} {Phys. Rev. A}\ }\textbf {\bibinfo {volume} {96}},\ \bibinfo
  {pages} {032322} (\bibinfo {year} {2017})}\BibitemShut {NoStop}%
\bibitem [{\citenamefont {Milazzo}\ \emph {et~al.}(2019)\citenamefont
  {Milazzo}, \citenamefont {Braun},\ and\ \citenamefont
  {Giraud}}]{Mil.Bra.Gir:19}%
  \BibitemOpen
  \bibfield  {author} {\bibinfo {author} {\bibfnamefont {N.}~\bibnamefont
  {Milazzo}}, \bibinfo {author} {\bibfnamefont {D.}~\bibnamefont {Braun}}, \
  and\ \bibinfo {author} {\bibfnamefont {O.}~\bibnamefont {Giraud}},\
  }\href@noop {} {\bibfield  {journal} {\bibinfo  {journal} {Phys. Rev. A}\
  }\textbf {\bibinfo {volume} {100}},\ \bibinfo {pages} {012328} (\bibinfo
  {year} {2019})}\BibitemShut {NoStop}%
\bibitem [{\citenamefont {Chryssomalakos}\ \emph {et~al.}(2018)\citenamefont
  {Chryssomalakos}, \citenamefont {Guzm\'{a}n-Gonz\'{a}lez},\ and\
  \citenamefont {Serrano-Ens\'{a}stiga}}]{Chr.Guz.Ser:18}%
  \BibitemOpen
  \bibfield  {author} {\bibinfo {author} {\bibfnamefont {C.}~\bibnamefont
  {Chryssomalakos}}, \bibinfo {author} {\bibfnamefont {E.}~\bibnamefont
  {Guzm\'{a}n-Gonz\'{a}lez}}, \ and\ \bibinfo {author} {\bibfnamefont
  {E.}~\bibnamefont {Serrano-Ens\'{a}stiga}},\ }\href@noop {} {\bibfield
  {journal} {\bibinfo  {journal} {J.{} Phys.{} A.}\ }\textbf {\bibinfo {volume}
  {51}},\ \bibinfo {pages} {165202} (\bibinfo {year} {2018})}\BibitemShut
  {NoStop}%
\bibitem [{\citenamefont {Bacry}(1974)}]{Bac:74}%
  \BibitemOpen
  \bibfield  {author} {\bibinfo {author} {\bibfnamefont {H.}~\bibnamefont
  {Bacry}},\ }\href@noop {} {\bibfield  {journal} {\bibinfo  {journal} {J.
  Math. Phys.}\ }\textbf {\bibinfo {volume} {15}},\ \bibinfo {pages} {1686}
  (\bibinfo {year} {1974})}\BibitemShut {NoStop}%
\bibitem [{\citenamefont {Varshalovich}\ \emph {et~al.}(1988)\citenamefont
  {Varshalovich}, \citenamefont {Moskalev},\ and\ \citenamefont
  {Khersonskii}}]{Var.Mos.Khe:88}%
  \BibitemOpen
  \bibfield  {author} {\bibinfo {author} {\bibfnamefont {D.}~\bibnamefont
  {Varshalovich}}, \bibinfo {author} {\bibfnamefont {A.}~\bibnamefont
  {Moskalev}}, \ and\ \bibinfo {author} {\bibfnamefont {V.}~\bibnamefont
  {Khersonskii}},\ }\href@noop {} {\emph {\bibinfo {title} {Quantum {T}heory of
  {A}ngular {M}omentum}}}\ (\bibinfo  {publisher} {World Scientific},\ \bibinfo
  {year} {1988})\BibitemShut {NoStop}%
\bibitem [{\citenamefont {Penrose}\ and\ \citenamefont
  {Rindler}(1990)}]{Pen.Rin:90}%
  \BibitemOpen
  \bibfield  {author} {\bibinfo {author} {\bibfnamefont {R.}~\bibnamefont
  {Penrose}}\ and\ \bibinfo {author} {\bibfnamefont {W.}~\bibnamefont
  {Rindler}},\ }\href@noop {} {\emph {\bibinfo {title} {Spinors and
  {S}pace-time, Vol.{} 1}}}\ (\bibinfo  {publisher} {Cambridge University
  Press},\ \bibinfo {year} {1990})\BibitemShut {NoStop}%
\bibitem [{\citenamefont {Landsberg}(2012)}]{Lan:12}%
  \BibitemOpen
  \bibfield  {author} {\bibinfo {author} {\bibfnamefont {J.~M.}\ \bibnamefont
  {Landsberg}},\ }\href@noop {} {\emph {\bibinfo {title} {Tensors: geometry and
  applications}}}\ (\bibinfo  {publisher} {American Mathematical Society},\
  \bibinfo {year} {2012})\BibitemShut {NoStop}%
\bibitem [{\citenamefont {Brink}\ and\ \citenamefont
  {Satchler}(1968)}]{BrinkSatchler68}%
  \BibitemOpen
  \bibfield  {author} {\bibinfo {author} {\bibfnamefont {D.}~\bibnamefont
  {Brink}}\ and\ \bibinfo {author} {\bibfnamefont {G.}~\bibnamefont
  {Satchler}},\ }\href@noop {} {\emph {\bibinfo {title} {Theory of Angular
  Momentum}}}\ (\bibinfo  {publisher} {Clarendon Press},\ \bibinfo {address}
  {Oxford},\ \bibinfo {year} {1968})\BibitemShut {NoStop}%
\bibitem [{\citenamefont {Koornwinder}(1981)}]{Koo:81}%
  \BibitemOpen
  \bibfield  {author} {\bibinfo {author} {\bibfnamefont {T.~H.}\ \bibnamefont
  {Koornwinder}},\ }\href@noop {} {\bibfield  {journal} {\bibinfo  {journal}
  {Nieuw Archief Voor Wiskunde}\ }\textbf {\bibinfo {volume} {29}},\ \bibinfo
  {pages} {140} (\bibinfo {year} {1981})}\BibitemShut {NoStop}%
\bibitem [{\citenamefont {Suma}\ \emph {et~al.}(2017)\citenamefont {Suma},
  \citenamefont {Sirsi}, \citenamefont {Hegde},\ and\ \citenamefont
  {Bharath}}]{Sum.Sir.Hed.Bha:17}%
  \BibitemOpen
  \bibfield  {author} {\bibinfo {author} {\bibfnamefont {S.}~\bibnamefont
  {Suma}}, \bibinfo {author} {\bibfnamefont {S.}~\bibnamefont {Sirsi}},
  \bibinfo {author} {\bibfnamefont {S.}~\bibnamefont {Hegde}}, \ and\ \bibinfo
  {author} {\bibfnamefont {K.}~\bibnamefont {Bharath}},\ }\href {\doibase
  10.1103/PhysRevA.96.022328} {\bibfield  {journal} {\bibinfo  {journal} {Phys.
  Rev. A}\ }\textbf {\bibinfo {volume} {96}},\ \bibinfo {pages} {022328}
  (\bibinfo {year} {2017})}\BibitemShut {NoStop}%
\bibitem [{\citenamefont {Fano}(1953)}]{Fan:53}%
  \BibitemOpen
  \bibfield  {author} {\bibinfo {author} {\bibfnamefont {U.}~\bibnamefont
  {Fano}},\ }\href@noop {} {\bibfield  {journal} {\bibinfo  {journal} {Phys.
  Rev.}\ }\textbf {\bibinfo {volume} {90}},\ \bibinfo {pages} {577} (\bibinfo
  {year} {1953})}\BibitemShut {NoStop}%
\bibitem [{\citenamefont {Chruscinski}\ and\ \citenamefont
  {Jamiolkowski}(2004)}]{Chr.Jam:04}%
  \BibitemOpen
  \bibfield  {author} {\bibinfo {author} {\bibfnamefont {D.}~\bibnamefont
  {Chruscinski}}\ and\ \bibinfo {author} {\bibfnamefont {A.}~\bibnamefont
  {Jamiolkowski}},\ }\href@noop {} {\emph {\bibinfo {title} {Geometric {P}hases
  in {C}lassical and {Q}uantum {M}echanics}}}\ (\bibinfo  {publisher}
  {Birkh\"auser},\ \bibinfo {year} {2004})\BibitemShut {NoStop}%
\bibitem [{\citenamefont {Agarwal}(2013)}]{Aga:13}%
  \BibitemOpen
  \bibfield  {author} {\bibinfo {author} {\bibfnamefont {G.~S.}\ \bibnamefont
  {Agarwal}},\ }\href@noop {} {\emph {\bibinfo {title} {Quantum Optics}}}\
  (\bibinfo  {publisher} {Cambridge University Press},\ \bibinfo {year}
  {2013})\BibitemShut {NoStop}%
\bibitem [{\citenamefont {D\"ur}\ \emph {et~al.}(2000)\citenamefont {D\"ur},
  \citenamefont {Vidal},\ and\ \citenamefont {Cirac}}]{Duer2000}%
  \BibitemOpen
  \bibfield  {author} {\bibinfo {author} {\bibfnamefont {W.}~\bibnamefont
  {D\"ur}}, \bibinfo {author} {\bibfnamefont {G.}~\bibnamefont {Vidal}}, \ and\
  \bibinfo {author} {\bibfnamefont {J.~I.}\ \bibnamefont {Cirac}},\ }\href@noop
  {} {\bibfield  {journal} {\bibinfo  {journal} {Phys.\ Rev.\ A}\ }\textbf
  {\bibinfo {volume} {62}},\ \bibinfo {pages} {062314} (\bibinfo {year}
  {2000})}\BibitemShut {NoStop}%
\bibitem [{\citenamefont {Giraud}\ \emph {et~al.}(2008)\citenamefont {Giraud},
  \citenamefont {Braun},\ and\ \citenamefont {Braun}}]{Giraud08}%
  \BibitemOpen
  \bibfield  {author} {\bibinfo {author} {\bibfnamefont {O.}~\bibnamefont
  {Giraud}}, \bibinfo {author} {\bibfnamefont {P.}~\bibnamefont {Braun}}, \
  and\ \bibinfo {author} {\bibfnamefont {D.}~\bibnamefont {Braun}},\
  }\href@noop {} {\bibfield  {journal} {\bibinfo  {journal} {Phys. Rev. A}\
  }\textbf {\bibinfo {volume} {78}},\ \bibinfo {eid} {042112} (\bibinfo {year}
  {2008})}\BibitemShut {NoStop}%
\bibitem [{\citenamefont {Biedenharn}\ and\ \citenamefont
  {Louck}(1981)}]{Bie.Lou:81}%
  \BibitemOpen
  \bibfield  {author} {\bibinfo {author} {\bibfnamefont {L.~C.}\ \bibnamefont
  {Biedenharn}}\ and\ \bibinfo {author} {\bibfnamefont {J.~D.}\ \bibnamefont
  {Louck}},\ }\href@noop {} {\emph {\bibinfo {title} {Angular momentum in
  quantum physics}}}\ (\bibinfo  {publisher} {Cambridge University Press},\
  \bibinfo {year} {1981})\BibitemShut {NoStop}%
\end{thebibliography}%
%
\end{document}